\documentclass[aps,pra,twocolumn,amsmath,amssymb,showpacs,floatfix,longbibliography]{revtex4-1}

\usepackage{amsmath}
\usepackage{txfonts}
\usepackage{microtype}
\usepackage{graphicx}
\usepackage{color}
\usepackage{ulem}
\usepackage{bm}

\usepackage{stmaryrd,tensind}
\usepackage{mathtools}
\usepackage{stmaryrd}

\newcommand{\Tr}{{\rm Tr}\,}
\newcommand{\Det}{{\rm Det}\,}
\newcommand{\comm}[1]{}

\usepackage{amsmath}
\usepackage{amsfonts}

\usepackage{amssymb}
\usepackage{amssymb}
\makeatletter
\newsavebox\myboxA
\newsavebox\myboxB
\newlength\mylenA

\newcommand*\xoverline[2][0.75]{%
    \sbox{\myboxA}{$\m@th#2$}%
    \setbox\myboxB\null% Phantom box
    \ht\myboxB=\ht\myboxA%
    \dp\myboxB=\dp\myboxA%
    \wd\myboxB=#1\wd\myboxA% Scale phantom
    \sbox\myboxB{$\m@th\overline{\copy\myboxB}$}%  Overlined phantom
    \setlength\mylenA{\the\wd\myboxA}%   calc width diff
    \addtolength\mylenA{-\the\wd\myboxB}%
    \ifdim\wd\myboxB<\wd\myboxA%
       \rlap{\hskip 0.5\mylenA\usebox\myboxB}{\usebox\myboxA}%
    \else
        \hskip -0.5\mylenA\rlap{\usebox\myboxA}{\hskip 0.5\mylenA\usebox\myboxB}%
    \fi}
\makeatother

\begin{document}

\author{Andre G. Campos}
\email{agontijo@mpi-hd.mpg.de}
\author{K. Z. Hatsagortsyan}
\author{C. H. Keitel}
\affiliation{Max Planck Institute for Nuclear Physics, Heidelberg 69117, Germany}

%\title{A novel method for the construction of Dirac spinors with well defined orbital angular momentum}

\title{ Construction of Dirac spinors for electron vortex beams in background electromagnetic fields}

\date{\today}

\begin{abstract}

Exact solutions of the Dirac equation, a system of four partial differential equations, are rare. The vast majority of them are for highly symmetric stationary systems. Moreover, only a handful of solutions for time dependent dynamics exists. Given the growing number of applications of high energy electron beams interacting with a variety of quantum systems in laser fields, novel methods for finding exact solutions to the Dirac equation are called for. We present a method for building up solutions to the Dirac equation employing a recently introduced approach for the description of spinorial fields and their driving electromagnetic fields in terms of geometric algebras. We illustrate the method by developing several stationary as well as non-stationary solutions of the Dirac equation with well defined orbital angular momentum along the electron's propagation direction.
The first set of solutions describe free electron beams in terms of Bessel functions as well as stationary solutions for both a homogeneous and an inhomogeneous magnetic field. The second set of solutions are new and involve a plane electromagnetic wave combined with a generally inhomogeneous longitudinal magnetic field. Moreover, the developed technique allows us to derive general physical properties of the dynamics in such field configurations, as well as  provides physical predictions on the self-consistent electromagnetic fields induced by the dynamics.

\end{abstract}

\maketitle

\section{Introduction}

Electron beams having a well defined orbital angular momentum (OAM) along its direction of propagation, also known as electron vortex beams \cite{Bliokh_2007}, are now ubiquitous in experimental physics \cite{Uchida_2010,Verbeeck_2010}, with applications ranging from fine probing of matter \cite{PhysRevLett.108.074802} to high energy particle collisions \cite{Ivanov_2011} and radiation processes \cite{PhysRevX.6.011006} (see also reviews \cite{BLIOKH20171,Lloyd_2017} and references therein). Recent proposals to employ relativistic twisted electron beams for Compton scattering \cite{Seipt_2014}, Mott scattering \cite{Serbo_2015}, and for radiative recombination \cite{Zaytsev_2017} convert the fundamental theoretical  problem of relativistic twisted electron beams \cite{Birula_2017,Barnett_2017} to a practical one. Self-consistent descriptions of the angular momentum properties of the electron by studying exact solutions to the Dirac equation were reported for cases in which the electrons are freely propagating \cite{bliokh2011relativistic,PhysRevLett.118.114801,PhysRevA.100.012108}, as well as interacting with a laser field \cite{hayrapetyan2014interaction,PhysRevLett.115.194801}, a homogeneous magnetic field \cite{van2017nonuniform} and more general field configurations \cite{PhysRevLett.121.043202}.

In this paper we provide a general method for constructing exact solutions to the Dirac equation for electron vortex beams using the newly developed approach of Relativistic Dynamic Inversion (RDI) \cite{campos2017analytic,PhysRevResearch.2.013051}, which encompasses several of the aforementioned studied solutions as well as giving novel ones, thus providing an unifying method for the construction of multiple solutions to the Dirac equation. Hence, each general solution provided by RDI gives rise to a whole family of solutions to the Dirac equation, each corresponding to an electromagnetic field of distinct symmetry.
%distinct electromagnetic field.

The starting point of RDI is Hestenes' formulation of the Dirac equation \cite{hestenes1967real,hestenes1973local} describing the electron in  terms of a set of conservation laws and constitutive relations for local observables, namely, for the spacial distribution and flow of charge density, mass, energy-momentum and angular momentum. As the definition of local suggests, the method is based on a fluid dynamics description of the Dirac theory, pioneered by Takabayasi \cite{takabayasi1957relativistic}. An important advantage of such point of view is that a clear classical interpretation can be given to the elements of the theory. For instance, the Dirac current is associated with the classical $4-$velocity of the electron, while its charge density coincides with the quantum mechanical probability distribution for normalizable spinors. Moreover, one assumes that the fluid carries spin density which is associated with the quantum mechanical expectation value of the Pauli spin operator. Given that the fluid streamlines correspond to the electron's trajectories in the given field, RDI relies on an intuition of how the electron in the quasiclassical approximation is expected to move in the desired field configuration for which the solutions to the Dirac equation are sought. In a nutshell, the geometric properties of the electron's trajectory in spacetime  (i.e., the fluid streamlines) come from the unique decomposition of the Dirac spinor as a product of Lorentz boosts and spatial rotations. Then, a set of equations for the vector potential in terms of the functions parameterising the boosts and rotations are derived directly from the Dirac equation. Unique solutions for those vector potential equations are found by imposing constraints on them based on an intuitive picture of the electromagnetic fields required to generate the prescribed electron's path.
Here we expand on the solutions presented in Ref. \cite{PhysRevResearch.2.013051}, shedding some light on RDI itself by presenting a full derivation of the technique directly from the Dirac equation, thus further clarifying important aspects of RDI that have been hitherto unexplored. In particular, the intimate relationship between the spinor parametrization and the appropriate form of the applied electromagnetic fields.

We briefly summarise our main results. In Sec. \ref{section1} we describe the RDI technique with emphasis on the derivation of all elements of the method, thus clarifying their relationship with the components of the standard Dirac four-component spinor. The plane wave solutions for the free electron is then described in detail, helping to highlight the motivation for the form of the Dirac spinor employed in the rest of the paper. In Sec. \ref{subsection2A} the general form of the Dirac spinor that give rise stationary solutions to the Dirac equation with OAM is constructed and its geometrical and physical meanings  thoroughly investigated. Then, starting with the particular case of zero longitudinal momentum, the solutions to spinor Bessel beam as well as for the homogeneous and inhomogeneous magnetic fields along the $z$ axis are derived in Secs. \ref{sec0}, \ref{sec1} and \ref{sec2}, respectively. The generalisation of those solutions to the case of nonzero longitudinal momentum is discussed in Sec. \ref{sec3}; it is noteworthy that in this case the form of the spin vector changes considerably. In Sec. \ref{subsection2B} the general form of the spinor describing solutions for the case of an electron interacting with a magnetic field along the $z$-axis and a plane wave field whose propagation direction is parallel to the magnetic field is derived. From such a general spinor, in Sec. \ref{sec4} a  closed form solution for an electron with OAM in a plane wave field (the so called Volkov-Bessel spinors \cite{hayrapetyan2014interaction}) is derived for the first time rather straightforwardly, thus highlighting the power of the RDI technique. In Sec. \ref{sec5} the solution for an electron in a combination of plane wave field whose propagation direction is parallel to a homogeneous magnetic field, known as the Redmond solution \cite{doi:10.1063/1.1704385,PhysRevA.27.2291}, is constructed. A  novel solution to the Dirac equation is given in Sec.~\ref{sec6} corresponding to a combination of plane wave field whose propagation direction is parallel to a inhomogeneous magnetic field. RDI allows one to show that for all solutions to the Dirac equation corresponding to the combination of plane electromagnetic waves and a magnetic field along the wave’s propagation direction with an arbitrary perpendicular profile, the Redmond solution stands out as being the only one corresponding to a relativistic coherent state \cite{doi:10.1002/andp.19834950102}. Moreover, as demonstrated in Ref. \cite{PhysRevResearch.2.013051}, the electromagnetic vector potential derived from RDI is guaranteed to obey Maxwell's equations. Furthermore, we show in Sec. \ref{subsection2C} that such nice feature is complemented with an important predictive property. We show that RDI allows to predict the self-consistent field when  the plane wave fields are added to the setup with the magnetic fields (the self-consistent field is created via disturbance of the current generating the initial magnetic field).
We then close in Sec. \ref{section3} with our conclusion and outlook for future work.

\section{Description of the method}\label{section1}

\subsection{Hestenes' formulation of Dirac theory}

The goal of the Hestenes' formulation of Dirac equation is to express it in terms of geometrical quantities and to provide a geometrical interpretation \cite{hestenes1967real,hestenes1973local,hestenes1975observables}.
In the standard representation the Dirac equation for an electron with charge $e$ and mass $m$ in an external electromagnetic field $A_\mu$ reads
\begin{align}\label{standardDirac}
\gamma^\mu\left(i\hbar\partial_\mu-eA_\mu\right)\psi=mc\psi,
\end{align}
where $\gamma^\mu=\boldsymbol{\eta}^{\mu,\nu}\gamma_\nu$, and $\boldsymbol{\eta}^{\mu,\nu}=\boldsymbol{\eta}_{\mu,\nu}$.
The Dirac matrices are defined as a set of irreducible matrices $\gamma_\mu$ which satisfy
\begin{align}\label{antiComm}
\gamma_\mu\gamma_\nu+\gamma_\nu\gamma_\mu=2\boldsymbol{\eta}_{\mu,\nu},
\end{align}
where the $\boldsymbol{\eta}_{\mu,\nu} = \mbox{diag}(1,-1,-1,-1)$ is the metric tensor of special relativity. Here we follow the convention that Greek indices take the values $0,1,2,3$ while Latin indices take the values $1,2,3$. Also, boldsymbol letters refers to $3$-dimensional vectors. The $\gamma_\mu$ over the complex numbers generates the complete algebra of $4\times4$ matrices.
The Dirac spinor $\psi$ is a column matrix with four complex components
\begin{align}\label{diracColum1}
\psi=\begin{pmatrix}
        \psi_1 \\ \psi_2 \\ \psi_3 \\ \psi_4
       \end{pmatrix}=\begin{pmatrix}
       r_0-ir_3 \\ r_2-ir_1 \\ s_3+is_0
        \\ s_1+is_2
       \end{pmatrix},
\end{align}
where the $r_\mu$'s and $s_\mu$'s are real functions of spacetime and $i=\sqrt{-1}$.
The representation (\ref{diracColum1}) in terms of the components $\psi_1,\psi_2,\psi_3,\psi_4$ presumes a specific representation of the Dirac matrices which is the standard representation
\begin{align}\label{diracmatrices}
\gamma_0=\begin{pmatrix}
    I & 0 \\
    0   &   -I
    \end{pmatrix},\quad\gamma_k=\begin{pmatrix}
    0 & -\sigma_k \\
    \sigma_k   &  0
    \end{pmatrix}
\end{align}
where $I$ is the $2\times2$ identity matrix and the $\sigma_k$ are the usual $2\times2$ Pauli matrices, that is, traceless Hermitian matrices satisfying
\begin{align}\label{PauliAlgebra}
\sigma_1\sigma_2\sigma_3=iI.
\end{align}

The new point of view is to interpret $\gamma_\mu$ as vectors of a spacetime reference frame. By definition the scalar product of these vectors are just the components $\boldsymbol{\eta}_{\mu,\nu}$ of the metric tensor. That is,
\begin{align}\label{dotProd}
\frac{1}{2}(\gamma_\mu\gamma_\nu+\gamma_\nu\gamma_\mu)=\gamma_\mu\cdot\gamma_\nu=\boldsymbol{\eta}_{\mu,\nu}.
\end{align}
They generate an associative algebra over the real numbers which has been called the  spacetime algebra by Hestenes \cite{hestenes1967real,hestenes1973local,hestenes1975observables}, since it provides a direct and complete algebraic characterisation of the geometric properties of Minkowski spacetime.

The further goal is to write $\psi$ in terms of spacetime algebra which is independent of the matrix representation in order to highlight its geometrical significance.
It goes as follows. In the spacetime algebra the quantities $\alpha_k=\gamma_k\gamma_0 \,(k=1,2,3)$ are to be interpreted as vectors relative to the inertial system specified by the time-like vector $\gamma_0$. The $\alpha_k$ generates an algebra over the real numbers which is isomorphic to the Pauli algebra. This fact is emphasised by writing
\begin{align}\label{STA}
\alpha_1\alpha_2\alpha_3=\gamma_0\gamma_1\gamma_2\gamma_3=\boldsymbol{i}=i\gamma_5,\quad\boldsymbol{i}^2=-1,
\end{align}
thus, $\boldsymbol{i}$ plays a similar role as $i$ does in the Pauli Algebra since it is a root of $-1$ while also obeying $\boldsymbol{i}\alpha_k=\alpha_k\boldsymbol{i}$ for $k=1,2,3$.
From the standard representation (\ref{diracmatrices}) it follows
\begin{align}\label{HestenesPauli}
&\alpha_k=\begin{pmatrix}
    0 & \sigma_k \\
    \sigma_k   &  0
    \end{pmatrix},\quad\gamma_5=\begin{pmatrix}
    0 & I \\
    I   &  0
    \end{pmatrix},\nonumber\\
    &\varepsilon_{ijk}\gamma^i\gamma^j=\boldsymbol{i}\alpha_k=\begin{pmatrix}
    i\sigma_k & 0 \\
   0  &  i\sigma_k
    \end{pmatrix},
\end{align}
where $\varepsilon_{ijk}$ is the Levi-Civita symbol.
Introducing the following basis in spinor space
\begin{align}\label{spinorBasis}
&u_1=\begin{pmatrix}
        1 \\ 0 \\ 0 \\ 0
       \end{pmatrix},\quad u_2=\begin{pmatrix}
        0 \\ 1 \\ 0 \\ 0
       \end{pmatrix},\quad u_3=\begin{pmatrix}
        0 \\ 0 \\ 1 \\ 0
       \end{pmatrix},
       %\nonumber\\       &
       \quad u_4=\begin{pmatrix}
        0 \\ 0 \\ 0 \\ 1
       \end{pmatrix},
\end{align}
%such that
and using the relations
\begin{align}
\gamma_0u_1&=u_1,\label{a}\\
\boldsymbol{i}\alpha_3&=\gamma_2\gamma_1u_1=iu_1,\label{b}\\
u_2&=-\boldsymbol{i}\alpha_2u_1,\quad u_3=\alpha_3u_1,\quad u_4=\alpha_1u_1\label{c},
\end{align}
one can represent a Dirac spinor $\psi$ in the form
\begin{align}\label{DiracS2}
\psi=\Psi u_1,
\end{align}
where $\Psi$ can be expressed as an element of the spacetime algebra by interpreting the $\gamma_\mu\gamma_0$ as vectors as follows
\begin{align}\label{HestenesG}
\Psi=&r_0+s_1\alpha_1+s_2\alpha_2+s_3\alpha_3
+\boldsymbol{i}\left(s_0-r_1\alpha_1-r_2\alpha_2-r_3\alpha_3\right).
\end{align}
This will help to make the geometrical significance of spinors explicit.
The Dirac equation can be expressed in terms of $\Psi$
\begin{align}
   \left(\hbar c \partial\!\!\!/ \Psi  \gamma_2\gamma_1 - c e A\!\!\!/ \Psi\right)\gamma_0u_1  =  m c^2 \Psi u_1,\label{Pre-Dirac-Hestenes}
   \end{align}
with $A\!\!\!/ = A^{\mu}  \gamma_{\mu}$, $\partial\!\!\!/ = \gamma^{\mu} \partial_{\mu}$. Since the spacetime vectors operating on $u_1$ generates a complete basis for the Dirac spinors as shown in (\ref{a})-(\ref{c}), Eq. (\ref{Pre-Dirac-Hestenes}) can be put in the form
\begin{align}
   \left(\hbar c \partial\!\!\!/ \Psi  \gamma_2\gamma_1 - c e A\!\!\!/ \Psi\right)\gamma_0  =  m c^2 \Psi,\label{Pre-Dirac-Hestenes2}
   \end{align}
or equivalently
\begin{align}
   \left(\hbar c \partial\!\!\!/ \Psi  \gamma_2\gamma_1 - c e A\!\!\!/ \Psi\right)  =  m c^2 \Psi\gamma_0.\label{Dirac-Hestenes-Eq4}
   \end{align}
The Eq. (\ref{Dirac-Hestenes-Eq4}) is fully consistent with the Dirac equation, and by using (\ref{DiracS2}), a solution of one equation can be  expressed as a solution of the other. Thus, it can fairly be called the Dirac equation in the language of spacetime algebra. It is a very general equation despite the explicitly appearance of $\gamma_0,\gamma_1$ and $\gamma_2$ in it, which are determined only within a proper Lorentz transformation. It cannot be overemphasised that the vectors $\gamma_0,\gamma_1$ and $\gamma_2$ appearing in (\ref{Dirac-Hestenes-Eq4}) need not be associated a priori with any coordinate frame. They are simply a set of arbitrarily chosen orthonormal vectors. Adoption of a coordinate frame with $\gamma_0$ as the time component is equivalent in the conventional theory to adopting a matrix representation for which $\gamma_0$ is Hermitian and the $\gamma_k$ are anti-Hermitian.

Up to this point the above explanation follows closely Hestenes' derivation presented in Ref. \cite{hestenes1975observables}. In order to highlight the origin of all parameters appearing in the factorisation of the matrix $\Psi$, we now propose a novel approach. We start by transforming $\Psi$ to a block diagonal form with the following unitary matrix $T=\frac{1}{\sqrt{2}}\left(1+\gamma_5\gamma_0\right)$. That is,
\begin{align}\label{diagonalPsi}
T\Psi T^\dagger&=\begin{pmatrix}
   Q & 0 \\
   0  &   \sigma_2 (Q^\dagger)^T\sigma_2
    \end{pmatrix},\nonumber\\
 Q&=r^0-i s^0 -i(r^k-is^k)\sigma_k,
\end{align}
where the superscript $T$ denotes the transposed. The determinant of $\Psi$
\begin{align*}
\Det[\Psi]=\left(r_\mu r^\mu-s_\mu s^\mu-2ir^\mu s_\mu\right)\left(r_\mu r^\mu-s_\mu s^\mu+2ir^\mu s_\mu\right),
\end{align*}
is a Lorentz invariant quantity thus being a scalar function. Hereinafter we will only consider cases in which $r_\mu r^\mu- s_\mu s^\mu$ and $r^\mu s_\mu$ are not simultaneously zero. Hence, $\Det[\Psi]\neq0$ which implies that $\Psi$ is an invertible matrix. Let us redefine $Q$ as $Q=sV$, with $s=\sqrt{r_\mu r^\mu-s_\mu s^\mu-2ir^\mu s_\mu}$ and $\Det[V]=1$. The function $s$ can be written as $s=\sqrt{\rho}e^{i\beta/2}$, where
\begin{align*}
\sqrt{\rho}&=\left[((r_\mu r^\mu-s_\mu s^\mu)^2+4(r^\mu s_\mu)^2\right]^{1/2},\\
\beta&=\arg(r_\mu r^\mu-s_\mu s^\mu\pm2ir^\mu s_\mu),
\end{align*}
where the ``$-$" sign is for $Q$, while the ``$+$" sign is for $\sigma_2 (Q^\dagger)^T\sigma_2$. Here we emphasise that the above definitions of $\beta$ and $\rho$ are novel. Finally, we have
\begin{align}\label{diagonalPsi2}
T\Psi T^\dagger&=\sqrt{\rho}\begin{pmatrix}
   e^{-i\beta/2}V & 0 \\
   0  &   e^{i\beta/2}\sigma_2 (V^\dagger)^T\sigma_2
    \end{pmatrix}.
 \end{align}
We then see that the $\beta$ parameter should be defined together with $V$, a fact recently recognised by Hestenes \cite{hestenes2019quantum}. Geometrically, it corresponds to a spacetime reflection which is modulated by the parameter $\beta$. By construction $V$ is a unimodular $2\times2$ complex matrix. It forms a group, the $SL(2,\mathbb{C})$, a complex $3$-dimensional manifold hence having $6$ degrees of freedom that are associated with the parameters of a boost and a rotation.
 In order to nail down the geometrical meaning of $\Psi$, it is instructive to check how this unitary transformation affects Eq. (\ref{DiracS2})
$$
T\psi=T\Psi T^\dagger Tu_1,\quad Tu_1=\frac{1}{\sqrt{2}}\begin{pmatrix}
        1 \\ 0 \\ 1 \\ 0
       \end{pmatrix}.
 $$
Given that $T$ is the unitary transformation connecting the Dirac representation with the Weyl representation (see the appendix to chapter 1 in Ref. \cite{thaller2013dirac}. A brief description is here provided in Appendix \ref{Spinors}) we see that $\Psi$ is the $4\times4$ double cover of the Lorentz group (see chapter 2 of Ref. \cite{ryder1996quantum}). Note that such representation of the Dirac spinor is equivalent to the so called ``polar spinors", see e.g. \cite{fabbri2019polar}. It then follows that
the matrix spinor $\Psi$ can be written as
\begin{align}\label{GeneralMatrixPsi}
\Psi &=\sqrt{\rho} \exp \left(\boldsymbol{i} \beta/2 \right)\mathcal{R},\\
\mathcal{R}&=T^\dagger\begin{pmatrix}
  V & 0 \\
   0  &  \sigma_2 (V^\dagger)^T\sigma_2
    \end{pmatrix}T,
\end{align}
where $\mathcal{R}$ is an unimodular $4\times4$ complex matrix corresponding to Lorentz transformations. Note that $T\boldsymbol{i}T^\dagger=\mbox{diag}(-i,-i,i,i)$. It is noteworthy that the scalar function $\beta$ (kwnon as the  Yvon-Takabayashi angle \cite{takabayasi1957relativistic,yvon1940equations}) is directly related to antiparticles (see, for instance, Ref. \cite{campos2020reply}).
Since $\mathcal{R}$ is an invertible square matrix, it can always be written in the  polar form:
\begin{equation}
\label{BU}
  \mathcal{R}=\mathcal{B}U
\end{equation}
 where $\mathcal{B}$ is a positive definite Hermitian matrix while $U$ is unitary. A more instructive decomposition can be made that follows from the positive-definiteness of $\mathcal{B}$, which is $\mathcal{R}=e^{X}U=Ue^{Y}$  where $Y=U^\dagger XU$ and $X$ is the unique Hermitian logarithm of $\mathcal{B}$. Moreover, given that $\Det[\mathcal{B}]=1$, we must have $\Tr[X]=0$. Such form is useful in the association of $\Psi$ with matrix Lie Groups.
 As explained in the Appendices \ref{Appendix1} and \ref{Spinors}, hereinafter we will identify $U$ with rotations and $\mathcal{B}$ with boosts. Thus $X$ can be written in the general form $X=-w(a^1\alpha_1+a^2\alpha_2+a^3\alpha_3)/2$, where $w$ us the electron's  rapidity and $(a_1,a_2,a_3)$ is a unit vector giving the direction of the boost.

\subsection{Geometrical interpretation}

\begin{figure}[h]
  \includegraphics[width=0.8\hsize]{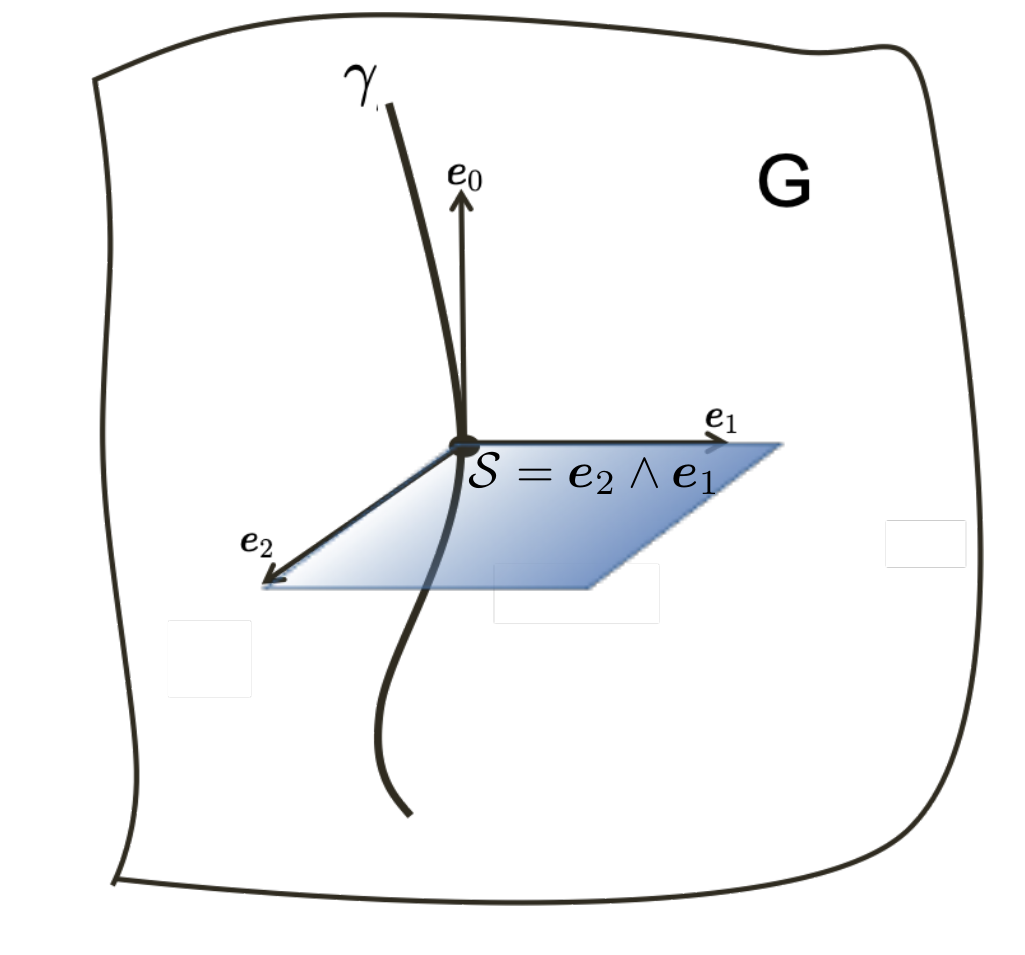}
  \caption{Illustration of the local orthonormal tetrad attached to the electron whose trajectory as seen by an observer in the laboratory frame of reference is given by the curve $\gamma$ on the manifold $G$. Three of the four orthonormal vectors composing the tetrad are shown. The time-like vector $\boldsymbol{e}_0$ is the tangent vector to the curve, thus describing the electron's velocity, while $\boldsymbol{e}_1$ and $\boldsymbol{e}_2$ are two space-like vectors whose geometrical meaning is elucidated in the text.
  The bivector $\mathcal{S}$ is depicted by the plane orthogonal to the electron's trajectory.}
     \label{ManifoldCurves}
\end{figure}

We are now ready to give a geometrical meaning to $\Psi$, which highlights the main idea behind RDI. At each spacetime point $x=(ct,\boldsymbol{r})$ the electron's rest frame $\{\gamma_\mu\}$ is connected to the laboratory frame $\{\boldsymbol{e}_\mu=\boldsymbol{e}_\mu(x)\}$ by the local Lorentz transformation $\mathcal{R}(x)$
\begin{align}\label{TetradFirst}
\boldsymbol{e}_\mu=\mathcal{R}\gamma_\mu\tilde{\mathcal{R}}=\mathcal{R}\gamma_\mu\gamma_0\mathcal{R}^\dagger\gamma_0,
\end{align}
with $\mathcal{R}\tilde{\mathcal{R}}=\tilde{\mathcal{R}}\mathcal{R}=1$. Whence, the matrix spinor (\ref{GeneralMatrixPsi}) determines the four mutually orthogonal vector fields
\begin{align}\label{electronFrame}
\Psi\gamma_\mu\tilde{\Psi}=\rho \boldsymbol{e}_\mu,
\end{align}
forming the orthonormal tetrad (OT)
\begin{align}\label{tetrads}
\rho(\boldsymbol{e}_0,\boldsymbol{e}_1,\boldsymbol{e}_2,\boldsymbol{e}_3)=(J,\rho \boldsymbol{e}_1,\rho \boldsymbol{e}_2,\rho \mathfrak{s}\!\!\!/),
\end{align}
that is attached to the electron. In other words, it exists at  every point along the electron's trajectory. The electron's $4-$velocity is given by the tangent vector field $J=\rho \boldsymbol{e}_0=\rho v\!\!\!/$, where $\rho v^\mu=(\psi^\dagger\psi,\psi^\dagger\alpha_1\psi,\psi^\dagger\alpha_2\psi,\psi^\dagger\alpha_3\psi)$ is the Dirac current obeying $\partial_\mu J^\mu=0$; it corresponds to the  velocity streamlines, i.e., timelike curves tangent to spacetime trajectories such as the one depicted in Fig. \ref{ManifoldCurves} describing the local flow of the  probability fluid. Moreover the spin density carried by the fluid is $\rho\mathfrak{s}\!\!\!/=\rho\boldsymbol{e}_3$, with $\rho\mathfrak{s}^\mu=(\psi^\dagger\boldsymbol{i}\psi,\psi^\dagger\boldsymbol{i}\alpha_1\psi,\psi^\dagger\boldsymbol{i}\alpha_2\psi,\psi^\dagger\boldsymbol{i}\alpha_3\psi)$ being the electron spin vector density whose direction is determined by the plane $\mathcal{S}=\mathcal{R}\gamma_2\gamma_1\tilde{\mathcal{R}}=\boldsymbol{e}_2\boldsymbol{e}_1=\boldsymbol{e}_2\wedge\boldsymbol{e}_1=\boldsymbol{i}\mathfrak{s}\!\!\!/\wedge v\!\!\!/$ which is also depicted in Fig. \ref{ManifoldCurves}. The spin and proper velocity obey the constraints: $v^\mu v_\mu=-\mathfrak{s}^\mu \mathfrak{s}_\mu=1$ and $v_\mu\mathfrak{s}^\mu=0$. The most important feature of a Lorentz transformation is that it leaves the metric tensor invariant, thus preserving the spacetime interval. As a consequence of this important property, the metric tensor in the laboratory frame $g_{\mu,\nu}$ is related to the metric tensor in the electron's frame $\boldsymbol{\eta}_{\mu,\nu}$ as
\begin{align}\label{MetricT}
g_{\mu,\nu}=\rho^2\boldsymbol{\eta}_{\mu,\nu},
\end{align}
which can be inferred from (\ref{electronFrame}).
Therefore, the Lorentz transformation (\ref{electronFrame}) acts as a conformal transformation, and we say that the metrics $g_{\mu,\nu}$ and $\boldsymbol{\eta}_{\mu,\nu}$ are conformally related. Thus we see that the scalar function $\rho$ acts as a dilatation. Conformal transformations commonly appears in the context of general relativity \cite{wald2007general}. Moreover, they are also connected with the transformation from an inertial coordinate system to another frame of reference with respect to which it is uniformly accelerated \cite{fulton1962physical}.
Furthermore, by identifying $J$ as the electron's current in the Dirac theory and $v\!\!\!/$ as its local $4-$velocity, $v_0\rho=J_0$ is also identified with the quantum mechanical probability distribution of finding the electron on a particular trajectory.
From Eq. (\ref{GeneralMatrixPsi}) we have, noting that $U\gamma_0=\gamma_0U$
\begin{align}
\Psi\gamma_0\tilde{\Psi}=\rho v\!\!\!/=\rho \mathcal{B}\gamma_0\mathcal{B}^{-1}=\rho \mathcal{B}^2\gamma_0,
\end{align}
since $\mathcal{B}^{-1}=\tilde{\mathcal{B}}=\gamma_0\mathcal{B}^\dagger\gamma_0=\gamma_0 \mathcal{B}\gamma_0$.
For the vector spin density, we have
\begin{align}
\Psi\gamma_3\tilde{\Psi}=\rho \mathfrak{s}\!\!\!/=\rho \mathcal{B}U\gamma_3U^\dagger \mathcal{B}^{-1}=\rho \mathcal{B} U\alpha_3 U^\dagger \mathcal{B}\gamma_0.
\end{align}
Once the matrix spinor is given, the next step is to find the electromagnetic fields that induce the motion of the electron encoded in $\Psi$. Formally, the vector potential can be written in terms of $ \Psi $ by inverting Eq. (\ref{Dirac-Hestenes-Eq4}) as
\begin{align}
 e A\!\!\!/= \hbar \partial\!\!\!/ \Psi  \gamma^2\gamma^1\Psi^{-1} -  m c \Psi \gamma^0\Psi^{-1},
   \label{GeneralA}
\end{align}
where
$$
\Psi^{-1}=\frac{\tilde{\Psi}}{\Psi\tilde{\Psi}},\quad \tilde{\Psi}=\gamma_0\Psi^\dagger\gamma_0,\quad \Psi\tilde{\Psi}=\rho e^{\boldsymbol{i}\beta}.
$$
Hence, the vector potential equation can be rewritten in a more illuminating form
\begin{align}
 e A\!\!\!/= \hbar \partial\!\!\!/ \Psi  \gamma^2\gamma^1\Psi^{-1} -  p\!\!\!/e^{-\boldsymbol{i}\beta},\quad  p\!\!\!/=mc v\!\!\!/,
   \label{GeneralA2}
\end{align}
which allows us to identify $p\!\!\!/$ with the  kinetic momentum.

The vector potential given by (\ref{GeneralA}) is required to obey the following constraints
\begin{align}\label{constraints}
&\Tr[e A\!\!\!/\Gamma_1]/4=0,\nonumber\\
&\Tr[e A\!\!\!/\Gamma_n]/4=0,\mbox{ for } 6\leq n\leq 16.
\end{align}
where
\begin{align*}
&\Gamma_1=\boldsymbol{1},\quad \Gamma_2=\gamma^0,\quad\Gamma_3=\gamma^1,\quad\Gamma_4=\gamma^2,\\
&\Gamma_5=\gamma^3,\quad\Gamma_6=\alpha_1,\quad\Gamma_7=\alpha_2,\quad\Gamma_8=\alpha_3,\nonumber\\
&\Gamma_9=\gamma^2\gamma^3,\quad \Gamma_{10}=\gamma^3\gamma^1,\quad\Gamma_{11}=\gamma^1\gamma^2,\\
&\Gamma_{12}=\gamma^1\gamma^2\gamma^3,\quad\Gamma_{13}=\gamma^0\gamma^2\gamma^3,\quad\Gamma_{14}=\gamma^0\gamma^3\gamma^1,\nonumber\\
&\Gamma_{15}=\gamma^0\gamma^1\gamma^2,\quad\Gamma_{16}=\gamma^5,
\end{align*}
that ensure, for instance, that the electron current obeys the continuity equation for the given matrix spinor parametrisation, i.e., $\partial_\mu J^\mu=0$. 

It is important to investigate further the bivector $\mathcal{S}$ since it will
play an important role in the equation for the vector potential. We first note that its components are
$$
\mathcal{S}=\gamma^0\gamma^i\mathfrak{s}^jv^k\varepsilon_{ijk}-\boldsymbol{i}\gamma^0\left(v_0\mathfrak{s}_k\gamma^k-\mathfrak{s}_0v_k\gamma^k\right),
$$
such that $\mathfrak{s}^jv^k\varepsilon_{ijk}=(\boldsymbol{\mathfrak{s}}\times\boldsymbol{v})^i$ corresponds to the $i-$th component of the cross product
between the spin and velocity vectors. If we define the ``matrix" vector $\vec{\alpha}=\gamma^0(\gamma^1,\gamma^2,\gamma^3)$ with $\vec{\alpha}\cdot\boldsymbol{v}=\gamma^0\gamma^kv_k$, we then have
\begin{align}\label{S}
\mathcal{S}=\vec{\alpha}\cdot\left(\boldsymbol{\mathfrak{s}}\times\boldsymbol{v}\right)-\boldsymbol{i}\vec{\alpha}\cdot\left(v_0\boldsymbol{\mathfrak{s}}-\mathfrak{s}_0\boldsymbol{v}\right).
\end{align}
The first thing we should note is that $\mathcal{S}$ has the same form as (\ref{groupPar}). Moreover, both of its terms are mutually orthogonal.

\subsection{Free particles}
Let us study the free particle solutions of Eq.  (\ref{Dirac-Hestenes-Eq4}). This will help us better understand and further develop the matrix spinor parametrisation. In this case, Eq. (\ref{GeneralA}) becomes
\begin{align}
 \hbar \partial\!\!\!/ \Psi  \gamma^2\gamma^1\Psi^{-1} =  m c \Psi \gamma^0\Psi^{-1}.
   \label{FreeCase}
\end{align}
From (\ref{GeneralMatrixPsi}) we have $\Psi\tilde{\Psi}=\rho e^{\boldsymbol{i}\beta}$,
hence
\begin{align}\label{PsiInverse}
\Psi^{-1}=\frac{\gamma_0\Psi^\dagger\gamma_0}{\rho e^{\boldsymbol{i}\beta}}=\frac{\gamma_0\mathcal{R}^\dagger\gamma_0 e^{-\boldsymbol{i}\beta/2}}{\sqrt{\rho}}.
\end{align}
Upon multiplying (\ref{FreeCase}) by $\gamma_0$ from the left we get
\begin{align}
 &\hbar \gamma_0\partial\!\!\!/ \Psi  \gamma^2\gamma^1\Psi^{-1} =  m c \gamma_0\Psi \gamma^0\Psi^{-1},\label{FreeCase2}\\
  &\gamma_0\partial\!\!\!/ =\frac{1}{c}\partial_t+\alpha_1\partial_x+\alpha_2\partial_y+\alpha_3\partial_z.
 \end{align}
Taking both $\rho$ and $\beta$ to be constants, Eq. (\ref{FreeCase2}) combined with (\ref{GeneralMatrixPsi}) leads to
\begin{align}
 \frac{\hbar}{2}\left(\gamma^0\gamma^\mu\Omega_\mu\right)\mathcal{R}  \gamma^2\gamma^1\tilde{\mathcal{R}} =  m c \gamma^0\mathcal{R}\gamma^0\tilde{\mathcal{R}}e^{-\boldsymbol{i}\beta}=mc\gamma^0v\!\!\!/e^{-\boldsymbol{i}\beta},
   \label{FreeCase3}
\end{align}
with $\Omega_\mu=2(\partial_\mu \mathcal{R})\tilde{\mathcal{R}}$. In what follows, it is more convenient to rewrite Eq. (\ref{FreeCase3}) as
\begin{align}
 \hbar\gamma^0\gamma^\mu\left(\partial_\mu\mathcal{R} \gamma^2\gamma^1\tilde{\mathcal{R}}\right) e^{\boldsymbol{i}\beta} =\gamma^0mc v\!\!\!/.
   \label{FreeCase4}
\end{align}
In order to proceed, let us take a closer look at $\mathcal{R}$. Noting that $\mathcal{B}U$ must be independent of coordinates in the solution of the free particle Dirac equation in terms of plane waves, from the right hand side of (\ref{FreeCase4})  we infer that $\mathcal{R}$ must be replaced by $\mathcal{R}e^{-\gamma^2\gamma^1\frac{p_\mu x^\mu}{\hbar}}$. That is
\begin{align}
\partial_\mu\mathcal{R}&\rightarrow\partial_\mu(\mathcal{R}e^{-\gamma^2\gamma^1\frac{p_\mu x^\mu}{\hbar}})\nonumber\\
&=\mathcal{R}e^{-\gamma^2\gamma^1\frac{p_\mu x^\mu}{\hbar}}(-\gamma^2\gamma^1\frac{p_\mu}{\hbar})\label{PhaseEq}
\end{align}
where $e^{-\gamma^2\gamma^1\frac{p_\mu x^\mu}{\hbar}}$ is a counterclockwise rotation round the $z$ axis by the Lorentz invariant angle $\frac{p_\mu x^\mu}{\hbar}$. Such rotation persists even in the electron's rest frame since there the electron still rotates by an angle $mc^2t/\hbar$.  Moreover, from the matrix form of the Dirac equation (\ref{GeneralMatrixPsi}), it is clear that such a phase factor must always be present in the matrix spinor parametrisation if the Dirac equation is to be satisfied. In fact, it can be shown that $\Psi\gamma_2\gamma_1\rightarrow i\psi$ which highlights the geometric meaning of the $i$ that appears in the Dirac equation. Hence, it is from the rotation matrix $e^{-\gamma^2\gamma^1\frac{p_\mu x^\mu}{\hbar}}$ that spin enters into the Dirac theory of the electron \cite{hestenes1973local}. In fact, given the importance of such term, Eq. (\ref{GeneralMatrixPsi}) will be amended as follows
\begin{align}\label{GeneralMatrixPsi2}
 \Psi =   \sqrt{\rho} \exp \left(\boldsymbol{i} \beta/2 \right)\mathcal{R}e^{-\gamma^2\gamma^1\frac{(Et-\mathbf{p}\cdot\mathbf{x}+\Phi)}{\hbar}},
\end{align}
where $\Phi$ is an arbitrary function of space and time whose presence changes Eq. (\ref{PhaseEq}) to
\begin{align}\label{PhaseEq2}
\partial_\mu\left(\mathcal{R}e^{-\gamma^2\gamma^1\frac{(Et-\mathbf{p}\cdot\mathbf{x}+\Phi)}{\hbar}}\right)=
\mathcal{R}e^{-\gamma^2\gamma^1\frac{(Et-\mathbf{p}\cdot\mathbf{x}+\Phi)}{\hbar}}\left[-\gamma^2\gamma^1\frac{(p_\mu+\partial_\mu\Phi)}{\hbar}\right].
\end{align}
Thus ensuring the  gauge invariance of the Hestenes-Dirac equation. Mathematically, the above amendment is due to the fact that the symmetry group underlying the Dirac theory is the Poincar\'e group, which is the Lorentz group augmented by the group of translations in spacetime generated by the momentum operator $p_\mu$ (see, for instance, section 2.7 of \cite{ryder1996quantum} for more details).

Let us rewrite Eq. (\ref{FreeCase4}) in the form
\begin{align}\label{FreeCase5}
p\!\!\!/ e^{\boldsymbol{i}\beta} =mc v\!\!\!/,
 \end{align}
from which we might be tempted to make the claim: for $\beta=0$ which corresponds to $r_\mu r^\mu-s_\mu s^\mu>0$ and $r^\mu s_\mu=0$ ($e^{\boldsymbol{i}\beta}=1$) the solution (\ref{FreeCase5}) describes particles, while for $\beta=\pi$ which corresponds to $r_\mu r^\mu-s_\mu s^\mu<0$ and $r^\mu s_\mu=0$ ($e^{\boldsymbol{i}\beta}=-1$) they describe antiparticles. However, this is not the case. The reason is that positive and negative energy states behave differently under boosts \cite{jehle1965relationship}. While a positive energy spinor of momentum $\mathbf{p}$ is obtained from a reference system in which it is at rest by a Lorentz boost to a new frame moving with velocity
$$
\mathbf{v}=-\frac{\mathbf{p}}{E_+}=-\frac{\mathbf{p}}{|E|},
$$
the negative energy spinor of the same momentum $\mathbf{p}$ transforms to the new frame which is moving with velocity
$$
\mathbf{v}'=-\frac{\mathbf{p}}{E_-}=\frac{\mathbf{p}}{|E|}.
$$
That is, both spinors must have opposite velocities. Hence, for a negative energy spinor, the following substitutions must be made in (\ref{FreeCase5}) \cite{hestenes1967real}
$$
\beta=\pi,\quad v\!\!\!/\rightarrow \gamma_0v\!\!\!/\gamma_0.
$$
From the above discussion, we see that a general free-particle solution $\Psi$ of the Dirac equation can be expanded in terms of plane waves
\begin{align}\label{GeneralFreePsi}
\Psi(x)&=\int\frac{d^3p}{(2\pi\hbar)^{3/2}}\sqrt{\frac{mc^2}{E}}\Big(\mathcal{R}e^{-\gamma^2\gamma^1\frac{(Et-\mathbf{p}\cdot\mathbf{x})}{\hbar}}\nonumber\\
&+e^{\frac{\pi}{2}\boldsymbol{i}}\gamma_0\mathcal{R}\gamma_0e^{\gamma^2\gamma^1\frac{(Et-\mathbf{p}\cdot\mathbf{x})}{\hbar}}\Big),
\end{align}
given that we can choose $\sqrt{\rho}=1$ without loss of generality for plane waves.

In order to have a better understanding of Eq. (\ref{GeneralFreePsi}), let us write down each of its components explicitly. For a boost with velocity $\mathbf{v}=-\mathbf{p}/E$ in an arbitrary direction, its matrix form is
\begin{align}\label{B2}
\mathcal{B}=\cosh\left(\frac{w}{2}\right)-\frac{\alpha_kv^k}{|\mathbf{v}|}\sinh\left(\frac{w}{2}\right).
\end{align}
Making the substitutions $\cosh(w/2)=\sqrt{\frac{E+mc^2}{2mc^2}}$, $\tanh(w/2)=-\frac{c|\mathbf{p}|}{E+mc^2}$ and
 $E=\sqrt{m^2c^4+\mathbf{p}^2c^2}$ in (\ref{B2}) leads to
\begin{align}\label{B3}
\mathcal{B}=\sqrt{\frac{E+mc^2}{2mc^2}}\left(1+\frac{c\alpha_kp^k}{E+mc^2}\right).
\end{align}
For a rotation with angle $\theta/2$ we have
\begin{align}\label{R2}
U=\cos(\theta/2)-\boldsymbol{i}\alpha_k\hat{\theta}^k\sin(\theta/2).
\end{align}
where the $\hat{\theta}^k$'s are the components of the unit vector along the rotation axis.
Now that we are done with the positive energy component of (\ref{GeneralFreePsi}), in case of the negative energy component we have
$$
\gamma_0\mathcal{B}U\gamma_0=\gamma_0\mathcal{B}\gamma_0U,
$$
since $\gamma_0U=U\gamma_0$. Thus
\begin{align*}
\gamma_0\mathcal{B}\gamma_0&=\gamma_0\sqrt{\frac{E+mc^2}{2mc^2}}\left(1+\frac{c\alpha_kp^k}{E+mc^2}\right)\gamma_0\\
&=\sqrt{\frac{E+mc^2}{2mc^2}}\left(1-\frac{c\alpha_kp^k}{E+mc^2}\right)=\mathcal{B}^-.
\end{align*}
Substituting the above equations in (\ref{GeneralFreePsi}) we end up with
\begin{align}\label{GeneralFreePsi2}
\Psi(x)&=\int\frac{d^3p}{(2\pi\hbar)^{3/2}}\sqrt{\frac{mc^2}{E}}\nonumber\\
&\times\Big(\mathcal{R}e^{-\gamma^2\gamma^1\frac{(Et-\mathbf{p}\cdot\mathbf{x})}{\hbar}}+e^{\frac{\pi}{2}\boldsymbol{i}}\mathcal{R}^-e^{\gamma^2\gamma^1\frac{(Et-\mathbf{p}\cdot\mathbf{x})}{\hbar}}\Big),
\end{align}
with $\mathcal{R}^-=\mathcal{B}^-U$.

The field free Dirac equation can be extracted from (\ref{GeneralFreePsi2}) and it is given by
\begin{align}\label{GenDiracS}
\psi(x)&=\int\frac{d^3p}{(2\pi\hbar)^{3/2}}\sqrt{\frac{mc^2}{E}}\nonumber\\
&\times\Big(\phi^{+}e^{-i\frac{(Et-\mathbf{p}\cdot\mathbf{x})}{\hbar}}+i\phi^{-}e^{i\frac{(Et-\mathbf{p}\cdot\mathbf{x})}{\hbar}}\Big),\nonumber\\
\phi^+&=\sqrt{\frac{E+mc^2}{2mc^2}}\begin{pmatrix}
       \xi^{(1/2)} \\ \frac{c\sigma_kp^k}{E+mc^2}\xi^{(1/2)}
       \end{pmatrix},\nonumber\\
        \phi^-&=\sqrt{\frac{E+mc^2}{2mc^2}}\begin{pmatrix}
      \frac{c\sigma_kp^k}{E+mc^2}\eta^{(1/2)}\\  -\eta^{(1/2)}
       \end{pmatrix},\nonumber\\
 \xi^{(1/2)}&=e^{-i\theta\frac{\sigma_k\hat{\theta}^k}{2}}\xi^{(1/2)}_0=\begin{pmatrix}
       \alpha \\ \eta
       \end{pmatrix},\nonumber\\
  \eta^{(1/2)}&=e^{-i\theta\frac{\sigma_k\hat{\theta}^k}{2}}\eta^{(1/2)}_0,
\end{align}
with the $2-$spinors corresponding to spin up $\xi^{(1/2)}_0=(1,0)^T$ and down $\xi^{(-1/2)}_0=(0,1)^T$, as well as the correspondingly conjugated spinors $\eta^{(-1/2)}_0=-i\sigma_2\xi^{(1/2)}_0=(0,1)^T$ and $\eta^{(1/2)}_0=-i\sigma_2\xi^{(-1/2)}_0=(-1,0)^T$.

We highlight the positive energy component of Eq.~(\ref{GenDiracS}) as a prototype for the further spinor parametrisation:
\begin{align}\label{FreeSP}
&\begin{pmatrix}
      e^{\frac{-iEt}{\hbar}}(E+mc^2)\alpha\\ e^{\frac{-iEt}{\hbar}}(E+mc^2)\eta \\ -ie^{\frac{-iEt}{\hbar}}\hbar c\left[\alpha\frac{\partial}{\partial z}+\eta(\frac{\partial}{\partial x}-i\frac{\partial}{\partial y})\right]\\ -ie^{\frac{-iEt}{\hbar}}\hbar c\left[-\eta\frac{\partial}{\partial z}+\alpha(\frac{\partial}{\partial x}+i\frac{\partial}{\partial y})\right]
       \end{pmatrix}\frac{e^{i\frac{\mathbf{p}\cdot\mathbf{x}}{\hbar}}}{\sqrt{2mc^2(E+mc^2)}}.
\end{align}
In fact, it should be emphasised that  all solutions presented here correspond to positive energy states only. Hence, in all cases we have
$r^\mu r_\mu-s^\mu s_\mu>0$ and $s^\mu r_\mu=0$.

\section{Applications of the method}\label{section2}

\subsection{Stationary solutions}\label{subsection2A}

For simplicity let us assume at the outset that any nontrivial dynamics happens on the $x$-$y$ plane only, so that the electron has constant momentum along the $z$ axis which we initially take to be zero. Moreover, let us consider the special case in which the electron undergoes circular motion on the plane. These assumptions will lead to the solutions for an electron with well defined OAM along its propagation direction. We then select a matrix spinor
\begin{equation}
\label{Psi_stationary}
 \Psi=\sqrt{\rho}\mathcal{B}Ue^{-\gamma^2\gamma^1\frac{(\varepsilon t+\Phi)}{\hbar}},
\end{equation}
having the following parametrisation
\begin{align}
U&=1\label{RotA},\\
\mathcal{B}&=e^{-\frac{w}{2}\left(\gamma^0\gamma^1\frac{-y}{\sqrt{x^2+y^2}}+\gamma^0\gamma^2\frac{x}{\sqrt{x^2+y^2}}\right)}\label{BAa},\\
&=e^{-\gamma^2\gamma^1\frac{1}{2}\tan^{-1}\left(\frac{y}{x}\right)}e^{-\frac{w}{2}\alpha_2}e^{\gamma^2\gamma^1\frac{1}{2}\tan^{-1}\left(\frac{y}{x}\right)}\label{BAb},\\
\Phi&=\frac{\hbar M}{2}\tan^{-1}\left(\frac{y}{x}\right),\label{PhaseR}\\
\frac{w}{2}&=\tanh^{-1}\left(\frac{B}{2(mc^2+\varepsilon)}\frac{d\ln(f(\lambda))}{d\lambda}\right)\label{rapidity},\\
\sqrt{\rho}&=\frac{(mc^2+\varepsilon)\lambda^{M/2}f(\lambda)H(\lambda)}{B\cosh(w/2)}\label{density},
\end{align}
where $\lambda=\frac{B \sqrt{x^2+y^2 }}{2 c \hbar }$, $f$ and $H$ are arbitrary real functions while $B$ is a positive real number with dimensions of energy to be determined later and the integer $M\geq0$ is the orbital angular momentum quantum number. It is noteworthy that $e^{-\gamma^2\gamma^1\frac{\varepsilon t+\Phi}{\hbar}}$ indicates the sense of rotation of the spin, which is counterclockwise. Moreover, It must be emphasised that the constant $B$ has  a priori no physical meaning, being just a convenient way to get the units right; its physical meaning will come from the vector potentials derived from the matrix spinor. Furthermore, the above spinor parameterization only ensures that the electron's velocity is along the azimuthal direction on the plane, and do not yet fully specify the shape of the trajectory, which could as well be elliptical. The special choice of circular trajectories will require further constraints on the form of the vector potential as we show bellow.

\begin{figure}[h]
  \includegraphics[width=0.7\hsize]{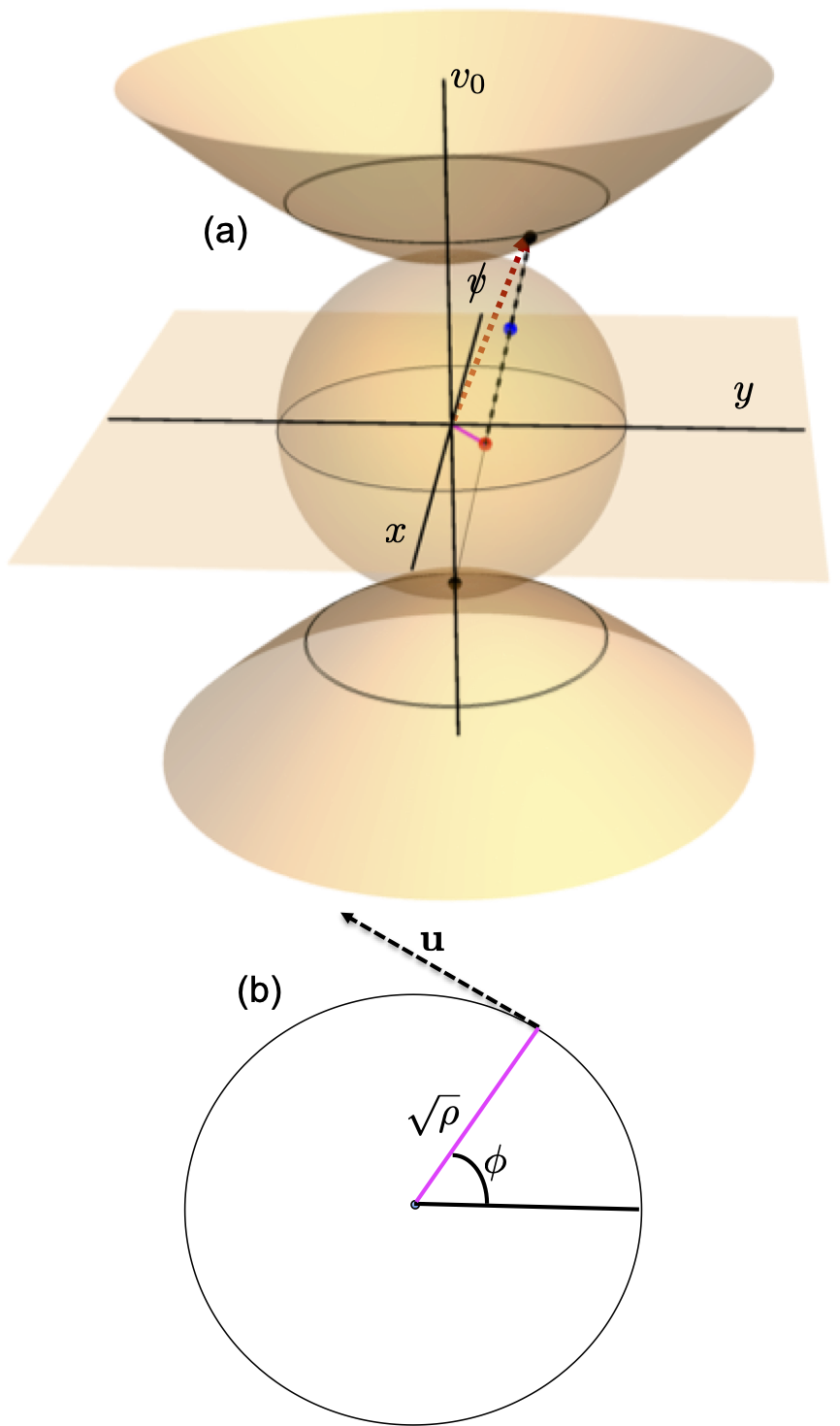}
  \caption{(Color Online). Geometrical meaning of the boost matrix Eq.~(\ref{BAa}) and velocity Eq.~(\ref{VelEqxy}) for a space of $2+1$ dimensions. (a) The upper sheet of the hyperboloid is mapped onto the plane passing through the origin by a stereographic projection from the south pole on the lower sheet of the hyperboloid. The sphere is the intersection between the plane and the light-cone with vertex on the particle's trajectory. The disk formed by the intersection between the plane and the sphere corresponds to the Poincar\'e disk model. (b) The circular trajectory on the upper sheet of the hyperboloid is mapped onto the origin centred circle of radius $\sqrt{\rho}$ and tangent vector $\mathbf{u}$. In the Figure, the centre of the sphere corresponds to the electron's rest frame $\{\gamma_\mu\}$ while the points on the upper sheet of the hyperboloid corresponds to the observer in the laboratory frame.}
     \label{Stereo}
\end{figure}

Let us now get a better understanding of the boost $\mathcal{B}$ whose matrix form is
\begin{align}
\mathcal{B}=u_0\left(1-\gamma_1\gamma_0 u^1-\gamma_2\gamma_0 u^2\right)
\end{align}
with
\begin{align*}
u_0&=\frac{1}{\sqrt{1-\frac{B^2(d\ln(f(\lambda))/d\lambda)^2}{4(mc^2+\varepsilon)^2}}},\\
\mathbf{u}&=\left(-\frac{y}{\sqrt{x^2+y^2}},\frac{x}{\sqrt{x^2+y^2}}\right)\frac{B }{2(mc^2+\varepsilon)}\frac{d}{d\lambda}\ln(f(\lambda))\\
&=\hat{\phi}\frac{B }{2(mc^2+\varepsilon)}\frac{d\ln(f(\lambda))}{d\lambda}.
\end{align*}
Since $\mathcal{B}$ is a  quaternion, its geometrical meaning is depicted in Fig. \ref{Stereo}. The vector $\mathbf{u}$ corresponds to the coordinates of the point in which the plane intersects the line from the south pole of the sphere to the upper sheet of the hyperboloid shown in Fig. \ref{Stereo} (a). Moreover, in the centre of the sphere we have $u_0=1$, in the points within the sphere we have $u_0<1$ and in the points on the surface of the sphere we have $\frac{B }{2(mc^2+\varepsilon)}\frac{d}{d\lambda}\ln(f(\lambda))=1$ which projects onto the point at infinity on the Poincar\'e disk model. In Fig. \ref{Stereo} (b) we see that $\mathcal{B}$ induces a motion in the azimuthal direction on the $x-y$ plane which is depicted by the tangent to the circle of radius $\sqrt{\rho}$. This point becomes obvious from Eq. (\ref{BAb}).
From Eqs. (\ref{electronFrame}) and (\ref{tetrads}) we have the following
expressions for the spin and velocity vectors, respectively
\begin{align}
\mathfrak{s}\!\!\!/&=\gamma^3\label{SpinEq},\\
v\!\!\!/&=\frac{1+|\mathbf{u}|^2}{1-|\mathbf{u}|^2}\left[\gamma^0-\frac{2}{1+|\mathbf{u}|^2}\left(\gamma^1u_1+\gamma^2u_2\right)\right]\nonumber\\
&=v_0\left(\gamma^0-\gamma^1v_1-\gamma^2v_2\right).\label{VelEqxy}
\end{align}
The velocity (\ref{VelEqxy}) is depicted as the dotted vector in Fig. \ref{Stereo} (a) and is written in terms of the coordinates on the plane. It is simply the stereographic projection from the upper sheet of the hyperboloid onto the plane passing through the origin of the coordinate system. The Dirac spinor $\psi$ is extracted from $\Psi$ and is given by
\begin{align}\label{spinorHomoB_a}
\psi&=H\left(\lambda\right)
 e^{-it \varepsilon/\hbar}\left(e^{i\tan^{-1}\left(\frac{y}{x}\right)}\lambda\right)^{M/2}\nonumber\\
 &\times\begin{pmatrix}\frac{\left(mc^2+\varepsilon \right)}{B}
   \\ 0 \\  0 \\ -\frac{i\hbar c}{B}\left(\frac{\partial}{\partial x}+i\frac{\partial}{\partial y}\right) \end{pmatrix}f\left(\lambda\right),
\end{align}
which has the same form as Eq.~(\ref{FreeSP}) with $p_z=\theta=\eta=0$ and $\alpha=1$; later on we will show how our solutions change by making $p_z\neq0$.

The vector potential follows from Eq. (\ref{GeneralA2})
\begin{align}
eA_0&=\frac{\hbar}{2}\left(\frac{\vec{\nabla}\cdot(\rho\boldsymbol{\mathfrak{s}}\times\boldsymbol{v})}{\rho}\right)+P_0-p_0\label{VecPotMagA0G},\\
e\boldsymbol{A}&=\frac{\hbar}{2}\left[\frac{1}{\rho}\vec{\nabla}\times(\rho\{v_0\boldsymbol{\mathfrak{s}}-\mathfrak{s}_0\boldsymbol{v}\})\right]-\boldsymbol{P}-\boldsymbol{p}\label{VecPotMagAvG},\nonumber\\
\end{align}
where
\begin{align}
P_0&=-\frac{\hbar}{8c}\Tr[\boldsymbol{e}_2\cdot\partial\boldsymbol{e}_1/\partial t]=\frac{\varepsilon}{ c},\nonumber\\
P_k&=-\frac{\hbar}{8}\Tr[\boldsymbol{e}_2\cdot\partial_k\boldsymbol{e}_1]\nonumber\\
&=-\left(\frac{M}{2}-\sinh^2(w/2)\right)\frac{\partial}{\partial x^k}\arctan\left(\frac{y}{x}\right),
\end{align}
with $(x^1,x^2)=(x,y)$.
Note that $A_3=0$. The components of (\ref{S}) appear in the equations for the vector field, which are both perpendicular to $\boldsymbol{v}$ given that $\mathfrak{s}_0=0$.

It is striking to notice that despite the vector potential given by Eqs. (\ref{VecPotMagA0G}) and (\ref{VecPotMagAvG}) already solving the Dirac equation for the spinor (\ref{spinorHomoB_a}), the extra condition $eA_0=0$ must be imposed in order to have circular trajectories. Geometrically it will become clear when we look at some examples that this condition implies there will be no fluid flow along the radial direction. Physically it means that the only force that might be acting on the electron is perpendicular to its velocity and that the electron moves on stable circular orbits. We then see the close connection between the classical picture inherited from the fluid dynamics point of view and the underlying physics of the problem. 

From the equations for $\rho$ and $\boldsymbol{v}$ we get, upon imposing the condition $A_0=0$
\begin{align}\label{PotCond}
&\frac{d^2f(\lambda)}{d\lambda^2}-\frac{4(m^2c^4-\varepsilon^2)f(\lambda)}{B^2}\nonumber\\
&+\frac{df(\lambda)}{d\lambda}\left(\frac{M+1}{\lambda}+\frac{2}{H(\lambda)}\frac{dH(\lambda)}{d\lambda}\right)=0,
 \end{align}
which implies
 \begin{align}
eA_1&=\frac{B^2y}{4c^2\lambda}\frac{d}{d\lambda}\ln(H(\lambda))\label{VecPotA1i},\\
eA_2&=-\frac{B^2x}{4c^2\lambda}\frac{d}{d\lambda}\ln(H(\lambda))\label{VecPotA2i}.
\end{align}
The magnetic field then becomes
\begin{align}\label{MagFB}
e\boldsymbol{B}&=-\frac{B^2 }{4c^2\lambda\hbar}\frac{d}{d\lambda}\left(\lambda\frac{d}{d\lambda}\ln(H(\lambda))\right)\hat{z}.
\end{align}
Hence, the magnetic field is dependent only on the explicit form of the function $H(\lambda)$. Thus, as we will show bellow,  whether or not the electron is free depends entirely on the explicit form of the electron's probability density $\rho v_0$.
There is another important point to emphasise regarding the form of the vector potential. In polar coordinates, it becomes
$$
e\boldsymbol{A}=\hat{\phi}A_\phi(\lambda),\quad A_\phi(\lambda)=-\frac{B\hbar}{2c}\frac{d\ln(H(\lambda))}{d\lambda}.
$$
We see that the vector potential has the form of a vortex, which is a  direct consequence of the boost $\mathcal{B}$. In other words, all of the solutions presented here will necessarily carry orbital angular momentum.

\subsubsection{Free particle case}\label{sec0}

From Eq. (\ref{MagFB}) we see that $H(\lambda)=1$ leads to $\boldsymbol{B}=0$. It then follows that the components of the vector potential (\ref{VecPotA1i}) and (\ref{VecPotA2i}) are zero and the solution of Eq. (\ref{PotCond}) is
\begin{align}\label{FreeE}
f(\lambda)=\lambda^{-l}J_{l}\left(\frac{2\lambda\sqrt{\varepsilon^2-m^2c^4}}{B}\right),
\end{align}
where $J$ denotes the Bessel function of the first kind and the substitution $M=2l$ was made. Note that $\varepsilon> mc^2$. We then recover the OAM Spinor Bessel state solution of the free Dirac equation reported in Ref. \cite{bliokh2011relativistic}. This can be seen as follows; by noting that $\tan^{-1}(y/x)=\phi$ and $(x+iy)/\sqrt{x^2+y^2}=e^{i\phi}$ the explicitly form of the Dirac spinor (\ref{spinorHomoB_a}) becomes
\begin{align}\label{BesselB2}
\psi=e^{-\frac{i\varepsilon t}{\hbar}}\begin{pmatrix}
       \frac{\mathcal{N} \left(c^2 m+\varepsilon \right) e^{i l \phi} J_l\left(\frac{2\lambda \sqrt{\varepsilon ^2-c^4 m^2}}{B }\right)}{B} \\ 0 \\ 0
        \\ \frac{i \mathcal{N}\sqrt{\varepsilon ^2-c^4 m^2} e^{i (l+1)\phi}
   J_{l+1}\left(\frac{2\lambda \sqrt{\varepsilon ^2-c^4 m^2}}{B }\right)}{B}
       \end{pmatrix},
\end{align}
which can be make equal to Eq. 7 from Ref. \cite{bliokh2011relativistic} for $\theta_0=\pi/2$, $\alpha=1$ and $\beta=0$ (note that these are the parameters appearing in Eq. 7 of  \cite{bliokh2011relativistic}) if we make the following identifications in Eq. (\ref{BesselB2}): $\mathcal{N}=\frac{B}{\sqrt{2}\varepsilon  \sqrt{\frac{c^2 m}{\varepsilon }+1}}$ and $\sqrt{\varepsilon ^2-c^4 m^2}=p_{\perp0}c$, where $p_{\perp0}$ is the initial transversal momentum as defined in  \cite{bliokh2011relativistic}.

It is noteworthy that if instead of choosing $H(\lambda)=1$, we choose $H(\lambda)=\lambda^{\zeta}$, where $\zeta$ is an arbitrary real number, then by changing $l\rightarrow l+\zeta$ in (\ref{BesselB2}) we get another free particle solution to the Dirac equation, this time corresponding to the  shifted Bessel beam as described in Ref. \cite{bliokh2012electron}. The free particle solution with $p_z\neq0$ will be discussed at the end of this section.

\subsubsection{Homogeneous magnetic field}\label{sec1}

Our solution of Eq.~(\ref{spinorHomoB_a}) describes also the case of a  constant magnetic field.
By solving Eq. (\ref{MagFB}) for a constant field, with $M=2l$, we get
\begin{align}\label{ChoseF2}
H(\lambda)=e^{-\lambda^2},
\end{align}
from which the vector potential and magnetic field follows
\begin{align}
eA_0&=0,\\
eA_3&=0,\\
eA_1&=-\frac{y B^2 }{2c^2\hbar},\\
eA_2&=\frac{ x B^2}{2c^2\hbar },\\
e\boldsymbol{B}&=\frac{B^2 }{c^2\hbar }\hat{z}.
\end{align}
The function $f(\lambda)$ and energy eigenvalues in this case take the form
\begin{align}
f(\lambda)&=\mathcal{N}L_n^{l}\left(2\lambda^2\right)\label{f2},\\
\varepsilon&=\sqrt{m^2c^4+2B^2n}\label{eigenHomo},
\end{align}
where $L_n^{l}\left(2\lambda^2\right)$ are the generalised Laguerre polynomials and $\mathcal{N}$ is a normalisation constant,
which comes from the normalisation condition $2\pi\int_0^\infty J_0\lambda d\lambda=1$:
\begin{align*}
\mathcal{N}=\frac{B}{l! \sqrt{2^{2-l}\pi \varepsilon  \left(c^2
   m+\varepsilon \right)}}\sqrt{\frac{(l+n)!}{n!}}.
\end{align*}
Moreover, both the radial and the $z$ components of the electron's current are zero, and the averaged azimuthal component is
\begin{align}\label{AverageJHB}
\langle J_\phi\rangle=\frac{\sqrt{2}Bn}{\varepsilon},
\end{align}
while the average electron's density is
\begin{align}\label{averagerhoHB}
\langle\rho\rangle=\frac{mc^2}{\varepsilon}.
\end{align}

The obtained solution is in accordance with the one studied in Ref. \cite{van2017nonuniform} corresponding to the case in which
the eigenvalues (\ref{eigenHomo}) are independent of $l$. However, if instead we make the substitution $M=-2l$ in Eq.~(\ref{spinorHomoB_a}), with $l$ being a
positive integer, we arrive at
\begin{align}
f(\lambda)&=\mathcal{N}(-1)^l\left(2 \lambda ^2 \right)^l \, _1F_1\left(-n;l+1;2 \lambda ^2 \right),\label{f3}\\
&=\mathcal{N}(-1)^l\left(2 \lambda ^2 \right)^l\binom{n}{n+l}L_n^{l}\left(2 \lambda ^2 \right),\nonumber\\
\varepsilon&=\sqrt{m^2c^4+2B^2(l+n)},\label{eigenHomo2}
\end{align}
leading to a solution of an electron in a homogeneous magnetic field for which the states of different angular momentum are no longer degenerate. The positive (negative) value of $M$
corresponds to making a clockwise (counterclockwise) rotation round the $z$ axis. Therefore, if the rotation induced by (\ref{PhaseR}) is  opposite ($M=2l$) to the direction of the velocity $\mathbf{u}$, the eigenstates with different
orbital angular momentum are degenerate, whereas if the induced rotation is in the  same  ($M=-2l$) direction as $\mathbf{u}$ the eigenstates with different
orbital angular momentum are non-degenerate.

As previously mentioned,  no projection onto positive energy states needs to be done here, since all our solutions to the homogeneous magnetic field case, which coincide with the ones presented in Ref.  \cite{van2017nonuniform}, already corresponds to positive energy states  only.

\subsubsection{Inhomogeneous magnetic field}\label{sec2}

For a simple example of an inhomogeneous magnetic field we substitute
\begin{align}\label{ChoseF3}
H(\lambda)=e^{-\frac{\lambda}{2}},
 \end{align}
 in Eq. (\ref{PotCond}) from which we get as solution
the function
 \begin{align*}
f(\lambda)=\mathcal{N} e^{ -\frac{\lambda \left(\mathcal{E}-B\right)}{2
   B}}U\left(\frac{(M+1) \left(\mathcal{E}-B\right)}{2 \mathcal{E}},M+1,\frac{\lambda\mathcal{E}}{B}\right),
  \end{align*}
with $\mathcal{N}$ being a normalisation constant and $\mathcal{E}=\sqrt{B^2+16 c^4 m^2-16 \varepsilon ^2}$; $U$ is Tricomi's confluent Hypergeometrical function. The vector potential, magnetic field and current then vary radially, given by
\begin{align}
eA_0&=eA_3=0,\label{VecPot3}\\
eA_1&=-\frac{y B }{4c \sqrt{x^2+y^2}},\label{VecPot1}\\
eA_2&=\frac{ x B}{4c \sqrt{x^2+y^2}},\label{VecPot2}\\
e\boldsymbol{B}&=\frac{B }{4c \sqrt{x^2+y^2}}\hat{z},\label{MagF}\\
e\mu_0\boldsymbol{J}_e&=\frac{B}{4c}\left(-\frac{y}{(x^2+y^2)^{3/2}},\frac{x}{(x^2+y^2)^{3/2}},0\right)\nonumber\\
&=\frac{e\boldsymbol{A}}{x^2+y^2}\label{current},
\end{align}
as can be easily verified by plugging in (\ref{ChoseF3}) in (\ref{VecPotA1i}) ,  (\ref{VecPotA2i}) and (\ref{MagFB}).
The vortex structure of the current (\ref{current}) is not surprising given the chosen form of the boost $\mathcal{B}$. It is then advantageous to also write it in polar coordinates $ e\mu_0\boldsymbol{J}_e=\hat{\phi}\frac{B}{4cr^2},\quad r=\sqrt{x^2+y^2}$, which  resembles the current carried by a circular loop on the $x-y$ plane. It is also solenoidal, that is, $\vec{\nabla}\cdot\boldsymbol{J}_e=0$.

The hypergeometric function $U$ above will be a finite polynomial if the following holds
\begin{align}\label{eigenInHomo}
\varepsilon=\sqrt{m^2c^4+\frac{ n (n+M+1)B^2}{4(2 n+M+1)^2}},
\end{align}
where the integers $n\geq0$ and $M\geq0$ are the principal and angular momentum quantum numbers, respectively. With these conditions, we have
\begin{align}
f(\lambda)&=\mathcal{N}e^{-\frac{\lambda  (M+1)}{2 (2 n+M+1)}+\frac{\lambda}{2}}U\left(-n,1+M,\frac{\lambda  (M+1)}{2 n+M+1}\right)\label{FunctionF},\\
&=\mathcal{N}e^{-\frac{\lambda  (M+1)}{2 (2 n+M+1)}+\frac{\lambda}{2}}L_n^M\left(\frac{\lambda  (M+1)}{2 n+M+1}\right).\label{FunctionF2}
\end{align}
This solution for the inhomogeneous magnetic field concurs with one
discussed in Ref. \cite{hautot1972solutions}.

An important point concerns the orbital angular momentum quantum number which, in contrast to the homogeneous magnetic field case, still appears in the energy eigenvalues whether or not the sign of $M$ is changed.
Moreover, the normalisation condition $2\pi\int_0^\infty J_0\lambda d\lambda=1$ leads to the following normalisation constant
\begin{align*}
\mathcal{N}=\frac{B^2(M+1)^{M/2+1}(n!)^{1/2}}{2 \sqrt{2} c \hbar  \sqrt{2\pi  (M+2 n+1)^{M+3} (M+n)!
   \varepsilon\left(c^2 m+\varepsilon \right)}}.
\end{align*}
Both the radial and the $z$ components of the electron's current are zero, and the averaged azimuthal component is
\begin{align}\label{AverageJIB}
\langle J_\phi\rangle=\frac{Bn(1+n+M)}{(1+2n+M)^2\varepsilon},
\end{align}
while the average electron's density is
\begin{align}\label{averagerho}
\langle\rho\rangle=\frac{mc^2}{\varepsilon},
\end{align}
the same as (\ref{averagerhoHB}).

From the above examples for the magnetic fields we note that $\langle\rho\rangle\varepsilon/c=\langle p_0\rangle$, and since the vector field $\rho\boldsymbol{\mathfrak{s}}\times\boldsymbol{v}$ is along the radial direction, this confirms our geometrical interpretation for the condition $eA_0=0$. Also, the form of $\langle\rho\rangle$, which is the same for both homogeneous and inhomogeneous magnetic field cases, has an important physical significance which is derived from the identification of the electron streamlines with classical trajectories. Such connection was first stablished for the free Dirac electron in Ref. \cite{PhysRev.97.1712}, from which we take the equation
\begin{align}\label{ClassicalS}
\frac{d}{ds}x^\mu=\psi^\dagger\gamma^0\gamma^\mu\psi,\quad \frac{d}{ds}=v^\mu\partial_\mu.
\end{align}
Geometrically the definition of $\frac{d}{ds}$ is unambiguous, since it is the directional derivative along the streamlines, and is analogous with the proper time in classical dynamics. For  the zeroth component of the velocity in (\ref{ClassicalS}) ($\mu=0$) we see at once that $d(ct)/ds=v_0=\psi^\dagger\psi=J_0=\rho v_0$. Since we have from classical relativistic mechanics $d(ct)/ds=\varepsilon/mc^2$, it then follows that $\langle\rho\rangle=ds/d(ct)$. Such identification was already explored in Ref. \cite{cabrera2019operational}.

Incidentally, another way to get the free particle solution from the solutions with either the homogeneous or the inhomogeneous magnetic fields is to first consider their ground states, that is, we put $n=M=0$. Then, the normalisation becomes $\mathcal{N}=\frac{B}{2mc^2}$. In the limit of $B\rightarrow0$, we have $f(\lambda)=H(\lambda)=1$, $\varepsilon=mc^2$ and the wavefunction (\ref{spinorHomoB_a}) goes over to the spinor of a free particle at rest with spin up, thus highlighting the meaning of $B/mc$ as being proportional to the magnitude of the electron's velocity in the above magnetic field examples.

\subsubsection{Case with $p_z\neq0$}\label{sec3}

In this case the Dirac spinor (\ref{spinorHomoB_a}) gets modified to
\begin{align}\label{spinorHomoB_b}
\psi&=H\left(\lambda\right)
 e^{-i\frac{ t \varepsilon-zp_z}{\hbar}}\left(e^{i\tan^{-1}\left(\frac{ y}{ x}\right)}\lambda\right)^{M/2}\nonumber\\
 &\times\begin{pmatrix}\frac{\left(mc^2+\varepsilon \right)}{B}
   \\ 0 \\  \frac{cp_z}{B} \\ -\frac{i\hbar c}{B}\left(\frac{\partial}{\partial x}+i\frac{\partial}{\partial y}\right) \end{pmatrix}f\left(\lambda\right),
\end{align}
while Eq. (\ref{PotCond}) becomes
\begin{align}\label{PotCond_b}
&\frac{d^2f(\lambda)}{d\lambda^2}-\frac{4(m^2c^4+c^2p_z^2-\varepsilon^2)f(\lambda)}{B^2}\nonumber\\
&+\frac{df(\lambda)}{d\lambda}\left(\frac{M+1}{\lambda}+\frac{2}{H(\lambda)}\frac{dH(\lambda)}{d\lambda}\right)=0.
 \end{align}
 The vector potential components (\ref{VecPotA1i}), (\ref{VecPotA2i}) and the rotation matrix (\ref{RotA}) remain the same while the energy eigenvalues (\ref{eigenHomo}), (\ref{eigenHomo2}) and (\ref{eigenInHomo}) simply have $p_z^2c^2$ added under the square root. Moreover, the boost (\ref{BAa}) now becomes
 \begin{align}
&\mathcal{B}=\frac{\cosh(w/2)}{\sqrt{(c^2 m+\varepsilon )^2-c^2p_z^2}} \gamma^0\Bigg[(c^2 m+\varepsilon)\gamma^0\nonumber\\
&+\frac{B\dot{f}\left(\lambda\right)}{2 \sqrt{x^2+y^2}  f\left(\lambda\right)}\left(y\gamma^1-x\gamma^2\right)+cp_z\gamma^3\Bigg],\nonumber\\
&\cosh(w/2)= \frac{1}{\sqrt{1-\frac{B^2
   \dot{f}(\lambda )^2}{4 ((c^2 m+\varepsilon )^2-c^2p_z^2)f(\lambda)^2}}}.
\end{align}

The new expressions for the velocity and spin vectors are
\begin{align}
v\!\!\!/&=\gamma_0\left(\frac{1+\frac{c^2p_z^2}{(mc^2+\varepsilon)^2}}{1-\frac{c^2p_z^2}{(mc^2+\varepsilon)^2}}\right)\frac{1+\frac{\hbar^2c^2|\vec{\nabla}\ln(f(\lambda))|^2}{(mc^2+\varepsilon)^2+c^2p_z^2}}{1-\frac{\hbar^2c^2|\vec{\nabla}\ln(f(\lambda))|^2}{(mc^2+\varepsilon)^2-c^2p_z^2}}\nonumber\\
&+\frac{2\hbar c}{1-\frac{\hbar^2c^2|\vec{\nabla}\ln(f(\lambda))|^2}{(mc^2+\varepsilon)^2-c^2p_z^2}}\Bigg(\gamma_1\frac{\partial\ln(f(\lambda))}{\partial y}-\gamma_2\frac{\partial\ln(f(\lambda))}{\partial x}\nonumber\\
&+\gamma_3\frac{cp_z(mc^2+\varepsilon)}{(mc^2+\varepsilon)^2-c^2p_z^2}\Bigg)\frac{mc^2+\varepsilon}{(mc^2+\varepsilon)^2-c^2p_z^2},\label{VelEqpz}\\
\mathfrak{s}\!\!\!/&=v^3\gamma_0+\frac{cp_z}{mc^2+\varepsilon}\left(v^1\gamma_1+v^2\gamma_2\right)+\frac{1+\frac{c^2p_z^2}{(mc^2+\varepsilon)^2}}{1-\frac{c^2p_z^2}{(mc^2+\varepsilon)^2}}\nonumber\\
&\times\left(\frac{1-\frac{\hbar^2c^2|\vec{\nabla}\ln(f(\lambda))|^2}{(mc^2+\varepsilon)^2+c^2p_z^2}}{1-\frac{\hbar^2c^2|\vec{\nabla}\ln(f(\lambda))|^2}{(mc^2+\varepsilon)^2-c^2p_z^2}}\right)\gamma_3\label{SpinEqpz},\\
\end{align}
from which we note that, in contrast to the $p_z=0$ case, the spin undergoes a more intricate dynamics. There is a simple reason for this feature, which is the Lorentz invariant condition $\mathfrak{s}^\mu v_\mu=0$ that both the velocity and spin vectors
 must obey. Given that for $p_z=0$ we have $\mathfrak{s}_0=0$, so it must be the case that $\boldsymbol{v}\cdot\boldsymbol{\mathfrak{s}}=0$. Since now $\mathfrak{s}_0\neq0$ it follows that $\mathfrak{s}_0=\boldsymbol{v}\cdot\boldsymbol{\mathfrak{s}}/v_0$ and the vector part of the electron's velocity and spin are no longer orthogonal to each other.
Moreover, the electron's density now becomes
\begin{align}
\sqrt{\rho}&=\lambda^{M/2}f(\lambda)H(\lambda)\frac{\sqrt{(mc^2+\varepsilon)^2-c^2p_z^2}}{B\cosh(w/2)}.
\end{align}

It is instructive to see how the solution for the free particle case changes for $p_z\neq0$ in order to show that our method fully reproduces the more general OAM Bessel spinor solution from Ref. \cite{bliokh2011relativistic}. The spinor is modified as
\begin{align}\label{free2}
\psi&=\mathcal{N}\left(
\begin{array}{c}
 \frac{\left(c^2 m+\varepsilon \right) J_l(\chi ) e^{\frac{i (l \phi  \hbar +p_z z-t \varepsilon )}{\hbar }}}{B}
   \\
 0 \\
 \frac{c p_z J_l(\chi ) e^{\frac{i (l \phi  \hbar +p_z z-t \varepsilon )}{\hbar }}}{B} \\
 \frac{i J_{l+1}(\chi ) \sqrt{\varepsilon ^2-c^2 \left(c^2 m^2+p_z^2\right)} e^{\frac{i ((l+1) \phi  \hbar
   +p_z z-t \varepsilon )}{\hbar }}}{B} \\
\end{array}
\right),
\end{align}
where $\chi=2\lambda \sqrt{\varepsilon ^2-c^2 \left(c^2 m^2+p_z^2\right)}/B$.
By making the following identifications of the parameters $\alpha$, $\beta$, $\theta_0$ and $(p_{\parallel0},p_{\perp0})$ (these are the initial longitudinal and transversal momentum, respectively) appearing in Eq. 7 of \cite{bliokh2011relativistic} with the ones in Eq. (\ref{free2}): $p_z\rightarrow p\cos(\theta_0)=p_{\parallel0}$, $\sqrt{\varepsilon ^2-c^2 \left(c^2 m^2+p_z^2\right)}\rightarrow p\sin(\theta_0)=p_{\perp0}$, $\alpha=1$, $\beta=0$, $\mathcal{N}=\frac{B}{\varepsilon  \sqrt{\frac{c^2 m}{\varepsilon }+1}}$ and $p=\frac{\sqrt{\varepsilon ^2-c^4 m^2}}{c}$ we recover Eq. 7 of  \cite{bliokh2011relativistic} for the spin up case.

\subsection{Non-stationary solutions for plane wave fields propagating along the $z$ axis}\label{subsection2B}

\begin{figure}[h]
  \includegraphics[width=1.\hsize]{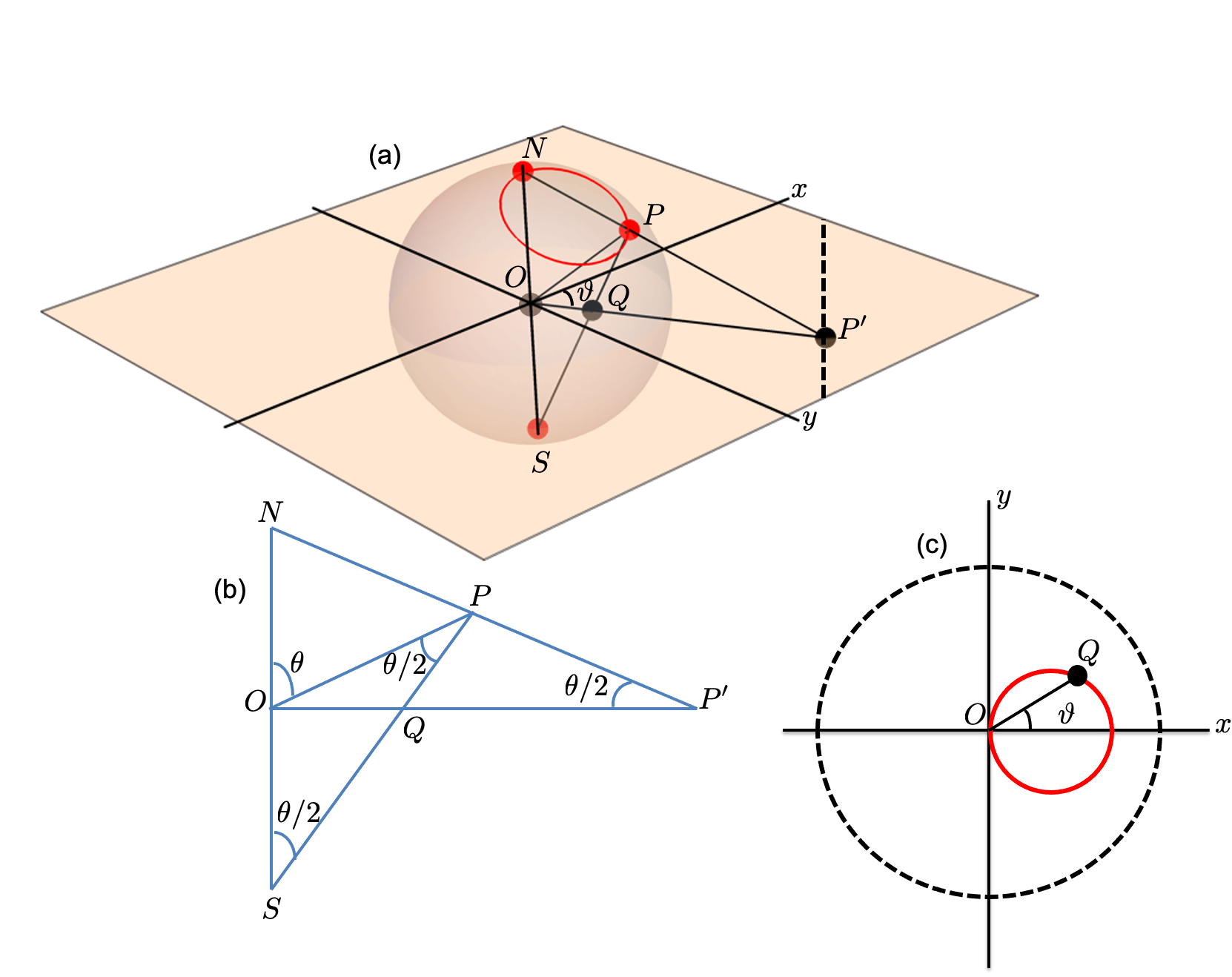}
  \caption{(Color Online). (a) The same sphere from Fig. \ref{Stereo} (a) is depicted here. The null rotation taking place on the surface of the sphere is given by the red full circle. A point $P$ on the circle is mapped by stereographic projection from the north pole $N$ onto the point $P'$ on the straight black dashed line on the plane. The point $P'$ is mapped onto the point $Q$ inside the sphere by a combination of a inversion and a complex conjugation on the coordinates of $P'$; the point $Q$ in turn corresponds to the same point $P$ under a stereographic projection from the south pole $S$ of the sphere. (b) The same triangles from the geometric construction in (a). (c) The dashed circle corresponds to the intersection between the unit sphere and the plane. The red full circle passing through $Q$ corresponds to the circle on the sphere mapped onto the plane by the Lorentz transformation $U_r\mathcal{B}_r$.}
     \label{Stereo2}
\end{figure}

To simplify the form of the equations, in this section we choose $p_z=0$. The generalisation to the non-zero momentum case is straightforward.
The starting point for the construction of the desired solution is the spinor of Eq.~(\ref{spinorHomoB_a}). The idea is to induce a so-called \textit{null rotation} (see pags. 28 and 29 of Ref. \cite{penrose1984spinors}) on the Riemann sphere shown in Fig. \ref{Stereo} (a). For instance, such a rotation is illustrated in Fig. \ref{Stereo2} (a) by the red full circle through the north pole of the sphere , which corresponds to the straight line on the $x-y$ plane passing through $P'$ by a stereographic projection from the north pole $N$ passing through $P$. Given that all the points outside of the sphere correspond to space-like trajectories, the point $P'$ is in turn mapped onto the point $Q$ by applying an inversion followed by a complex conjugation on the coordinates of $P'$. This transformation maps the line passing through $P'$ onto a circle tangent to the origin $O$ and passing through $Q$ (see Pag. 126 of Ref. \cite{needham1998visual}) which thus corresponds to the electron trajectory and is illustrated in Fig. \ref{Stereo2} (c). Moreover, from the triangles in  Fig. \ref{Stereo2} (b), we can get the following segment lengths
\begin{align*}
&SO=ON=OP=1,\quad SQ=\frac{1}{\cos(\theta/2)},\quad SP=2\cos(\theta/2),\\
&NP=2\sin(\theta/2),\quad NP'=\frac{1}{\sin(\theta/2)},\quad OQ=\tan(\theta/2),\\
&OP'=\cot(\theta/2).
\end{align*}
We are now ready to construct the null-rotation; it is defined by the following Lorentz transformation applied to $\Psi$ (see Appendix \ref{Spinors} for a detailed discussion)
\begin{align}
\Psi_T&=e^{k\!\!\!/\wedge \mathcal{A}\!\!\!/}\Psi e^{\gamma_2\gamma_1\Phi}=\left(1+k\!\!\!/\wedge \mathcal{A}\!\!\!/\right)\Psi e^{\gamma_2\gamma_1\Phi}\label{LorentzNull},\\
k\!\!\!/\wedge \mathcal{A}\!\!\!/&=\frac{c^2}{2\varepsilon\omega}\left(\dot{f_1}(\xi )\alpha_1+\dot{f_2}(\xi )\alpha_2-\dot{f_2}(\xi )\gamma_2\gamma_3+\dot{f_1}(\xi )\gamma_3\gamma_1\right)\nonumber,\\
&=\frac{c^2}{2\varepsilon\omega}\left[\vec{\alpha}\cdot\dot{\boldsymbol{f}}(\xi)-\boldsymbol{i}\vec{\alpha}\cdot(\dot{\boldsymbol{f}}(\xi)\times\hat{z})\right]\label{S2}
\end{align}
with $\xi=\omega(t-z/c)$, $\dot{f_i}(\xi )=df_i(\xi)/d\xi$ and $\dot{\boldsymbol{f}}(\xi)=(\dot{f_1}(\xi ),\dot{f_2}(\xi ),0)$. The translation of the electron on the $x-y$ plane is described by the vector $X_k=\frac{c^3}{2\varepsilon\omega^2}(f_1(\xi),f_2(\xi),0)$ while its velocity on the plane is $\mathcal{A}_\mu=\omega(0,\dot{X}_1(\xi)/c,\dot{X}_2(\xi)/c,0)$, which corresponds to the electron's classical trajectory on the laser field.
Moreover, the term applied to the right of $\Psi$ is a gauge transformation given by the function $
\Phi=-\frac{c^4}{2\varepsilon\omega^3\hbar}\int_0^\xi d\phi(\dot{f}_1(\phi)^2+\dot{f}_2(\phi)^2)$. The $\omega$ and $k_\mu=\frac{\omega}{c}(1,0,0,1)$ are the plane wave's frequency and wave vector, respectively. Note the  striking similarity between Eqs.~(\ref{S}) and (\ref{S2}), which is  not a coincidence. In fact, it is because $k\!\!\!/\wedge \mathcal{A}\!\!\!/$ plays a similar role as $\mathcal{S}$; it defines a plane tangent to the light-cone whose generator is the wave vector $k\!\!\!/$. Given that every null vector is orthogonal to itself, the bivector $k\!\!\!/\wedge \mathcal{A}\!\!\!/$ gives the laser field's propagation direction. The Lorentz transformation $e^{k\!\!\!/\wedge \mathcal{A}\!\!\!/}$, which has the same form as (\ref{groupPar}), can also be put in the polar form $e^{\,k\!\!\!/\wedge \mathcal{A}\!\!\!/}=U_r\mathcal{B}_r$ with the following boost and rotation matrices\\
\begin{align}\label{PolarLaser}
U_r&=e^{\theta\left(\cos(\vartheta)\gamma^1\gamma^3+\sin(\vartheta)\gamma^2\gamma^3\right)/2},\\
\mathcal{B}_r&=e^{-w(V_1\alpha^1+V_2\alpha^2+V_3\alpha^3)/2},\\
\cos(\vartheta)&=\frac{\dot{f}_1(\xi)}{\sqrt{\dot{f}_1(\xi)^2+\dot{f}_2(\xi)^2}},\\
\sin(\vartheta)&=\frac{\dot{f}_2(\xi)}{\sqrt{\dot{f}_1(\xi)^2+\dot{f}_2(\xi)^2}},\\
\vartheta&=\tan^{-1}\left(\dot{f}_2(\xi)/\dot{f}_1(\xi)\right),\\
\frac{\theta}{2}&=\tan^{-1}\left(c^2\frac{\sqrt{\dot{f}_1(\xi)^2+\dot{f}_2(\xi)^2}}{2\varepsilon\omega}\right),\\
\frac{V_1}{2}&=\cos\left(\frac{\theta}{2}\right)\cos(\vartheta),\\
\frac{V_2}{2}&=\cos\left(\frac{\theta}{2}\right)\sin(\vartheta),\\
\frac{V_3}{2}&=\sin\left(\frac{\theta}{2}\right),\\
w&=\tanh^{-1}\left(\frac{V_3}{2}\right).
\end{align}
Note that the velocity components (i.e., the components of the boost exponent) are simply the coordinates of the point $P$ on the sphere depicted in Fig. \ref{Stereo2} with respect to the south pole $S$. Moreover, the rotation matrix $U_r$ simply defines the direction of the velocity of the electron, which is tangent to the circles of latitude on the sphere. Furthermore, the gauge transformation can be written in terms of $\theta$ as
$$
\tan^2\left(\frac{\theta}{2}\right)=-\frac{\hbar\omega}{2\varepsilon}\frac{d\Phi}{d\xi},
$$
which is just the squared distance between the points $O$ and $Q$ shown in Fig. \ref{Stereo2}.

The general matrix spinor (\ref{LorentzNull}) can now be put in the form
\begin{align}\label{LorentzNull2}
\Psi_T=\sqrt{\rho}U_r\mathcal{B}_r\mathcal{B}e^{-\gamma_2\gamma_1\left(\frac{\varepsilon t}{\hbar}-\frac{M}{2}\arctan\left(\frac{y'}{x'}\right)-\Phi\right)},
\end{align}
whose application induces the following change of coordinates
\begin{align}\label{transfromedC}
 x'&= x+\frac{c^3f_1(\xi)}{\varepsilon\omega ^2 }\nonumber \\
 y'&= y+\frac{c^3f_2(\xi)}{\varepsilon\omega ^2},
\end{align}
with $\sqrt{\rho}$ given by Eq. (\ref{density}) in terms of the transformed coordinates (\ref{transfromedC}). Moreover, the velocity and spin vectors now become
\begin{align*}
\mathfrak{s}\!\!\!/_r&=-\gamma^0c^4\left(\frac{\dot{f}_1(\xi)^2+\dot{f}_2(\xi)^2}{2\varepsilon^2\omega^2}\right)+\gamma^1\frac{c^2\dot{f}_1(\xi)}{\varepsilon\omega}+\gamma^2\frac{c^2\dot{f}_2(\xi)}{\varepsilon\omega}\\
&+\gamma^3\left(1-c^4\frac{(\dot{f}_1(\xi)^2+\dot{f}_2(\xi)^2)}{2\varepsilon^2\omega^2}\right),\\
\boldsymbol{v}\!\!\!/_r&=\boldsymbol{v}\!\!\!/(x',y')-v_0(x',y')(\gamma_1\mathfrak{s}_r^1+\gamma_2\mathfrak{s}_r^2)\\
&-\left[v_0(x',y')\mathfrak{s}_r^0+v_1(x',y')\mathfrak{s}_r^1+v_2(x',y')\mathfrak{s}_r^2\right](\gamma_0+\gamma_3),
\end{align*}
where $v_\mu(x',y')$ are the components of the velocity (\ref{VelEqxy}) in terms of the transformed coordinates (\ref{transfromedC}). Again we see that the spin vector is no longer fixed in contrast with (\ref{SpinEq}).

 \subsubsection{General formula for the vector potential}

The exact expression for the vector potential, which generalises Eqs. (\ref{VecPotMagA0G}) and (\ref{VecPotMagAvG}), is derived from Eq. (\ref{GeneralA}) for the matrix spinor (\ref{LorentzNull2})
\begin{align}
eA_0^r&=\frac{\hbar}{2}\left(\frac{\vec{\nabla}'\cdot(\rho\boldsymbol{\mathfrak{s}}_r\times\boldsymbol{v}_r)}{\rho}\right)+P_0-mcv^0_r\label{VecPotMagA0L},\\
e\boldsymbol{A}^r&=\frac{\hbar}{2}\left[\frac{1}{\rho}\vec{\nabla}'\times(\rho\{v^0_r\boldsymbol{\mathfrak{s}}_r-\mathfrak{s}^0_r\boldsymbol{v}_r\})-\frac{1}{c\rho}\frac{\partial}{\partial t}(\rho\boldsymbol{\mathfrak{s}}_r\times\boldsymbol{v}_r)\right]\nonumber\\
&-\boldsymbol{P}-mc\boldsymbol{v}_r\label{VecPotMagAvL},\\
P_0&=-\frac{\hbar}{8c}\Tr[\boldsymbol{e}_2\cdot\partial\boldsymbol{e}_1/\partial t],\nonumber\\
P_k&=-\frac{\hbar}{8}\Tr[\boldsymbol{e}_2\cdot\partial_k\boldsymbol{e}_1],\nonumber\\
\nonumber
\end{align}
with $(x^1,x^2,x^3)=(x',y',z)$ and $\vec{\nabla}'=(\partial/\partial x',\partial,\partial y',\partial/\partial z)$.

\subsubsection{Transformed vector potential}\label{subsection2C1}

Let us investigate how the vector potentials (\ref{VecPotA1i}) and (\ref{VecPotA2i}) for the stationary solution cases transforms under the action of the Lorentz transformation $\Phi^r=e^{\,k\!\!\!/\wedge \mathcal{A}\!\!\!/}$ followed by the
coordinate transformation (\ref{transfromedC})
\begin{align}
&\Phi^r\left(\gamma^1\frac{B^2y}{4c^2\lambda}\frac{d}{d\lambda}\ln(H(\lambda))-\gamma^2\frac{B^2x}{4c^2\lambda}\frac{d}{d\lambda}\ln(H(\lambda))\right)\tilde{\Phi}^r\nonumber\\
&=\gamma^0eA_0^R+\gamma^1\frac{B^2y'}{4c^2\lambda'}\frac{d}{d\lambda'}\ln(H(\lambda'))-\gamma^2\frac{B^2x'}{4c^2\lambda'}\frac{d}{d\lambda'}\ln(H(\lambda'))\nonumber\\
&+eA_3^R\gamma^3=\gamma^\mu eA_\mu^R,\nonumber\\
A_0^R&=\frac{B^2  \left(y' \dot{f}_1(\xi )-x \dot{f}_2(\xi )\right)}{4 \lambda'  \omega
   \epsilon}\frac{d\ln(H(\lambda' ))}{d\lambda'},\quad A_3^R=A_0^R.\label{RFields}
\end{align}
where $\lambda'=\frac{B \sqrt{x'^2+y'^2} }{2 c \hbar }$. We conveniently labeled them as ``radiation fields" with subscript R. The reason for this is explained in Sec. \ref{subsection2C}.

There are again three different cases to consider which are discussed in the next subsection. Before proceeding, it is noteworthy that for the free particle case the vector potential can only be calculated from Eqs. (\ref{VecPotMagA0L}) and (\ref{VecPotMagAvL}). However, for the magnetic field cases, the solution to the general equation for the vector potential can be constructed by simply adding the free particle vector potential to the radiation potentials (\ref{RFields}). This conforms with the fact that the vector potential enters linearly into the Dirac equation.

\subsubsection{Free particle in a laser field}\label{sec4}

By making $H(\lambda')=1$ and $f(\lambda')$ equal to (\ref{FreeE}) in (\ref{VecPotMagA0L}) and (\ref{VecPotMagAvL}) one arrives at
\begin{align}\label{VecPotTD1}
eA_0^r&=0,\nonumber\\
 eA_1^r&=\frac{c  \dot{f}_1(\xi)}{ \omega }=\frac{2\varepsilon}{c}\mathcal{A}_1,\nonumber\\
 eA_2^r&=\frac{c  \dot{f}_2(\xi)}{ \omega }=\frac{2\varepsilon}{c}\mathcal{A}_2,\nonumber\\
 eA_3^r&=0.
\end{align}
From the OAM spinor Bessel beam solution of the Dirac equation,
we recover a generalization of the OAM Volkov-Bessel state given in Ref. \cite{hayrapetyan2014interaction}, as we  show below. After the Lorentz transformation (\ref{LorentzNull}) now applied to the matrix spinor representation of the OAM Bessel spinor (\ref{BesselB2}), the transformed wavefunction becomes
  \begin{widetext}
\begin{align}\label{BesselW}
\psi_T=F(x',y',\xi)\left(
\begin{array}{c}
  2 \left(c^2 m+\varepsilon \right)
   e^{il\phi}J_l\left(\frac{2\lambda'\sqrt{\left(\varepsilon ^2-c^4 m^2\right)}}{B
   }\right)-\frac{c \sqrt{\varepsilon ^2-c^4 m^2} e^{i (l+1)\phi}
   \left(\dot{f}_2(\xi)+i \dot{f}_1(\xi)\right)
   J_{l+1}\left(\frac{2\lambda'\sqrt{\left(\varepsilon ^2-c^4 m^2\right)}}{B }\right)}{\varepsilon \omega/c } \\
 \frac{c \left(c^2 m+\varepsilon \right) \left(\dot{f}_1(\xi)+i \dot{f}_2(\xi)\right)
   e^{il\phi}J_l\left(\frac{2\lambda'\sqrt{ \left(\varepsilon ^2-c^4 m^2\right)}}{B
   }\right)}{ \varepsilon \omega/c } \\
 -\frac{c \sqrt{\varepsilon ^2-c^4 m^2}
   \left(\dot{f}_2(\xi)+i \dot{f}_1(\xi)\right)e^{i(l+1) \phi}
   J_{l+1}\left(\frac{2\lambda'\sqrt{ \left(\varepsilon ^2-c^4 m^2\right)}}{B}\right)}{ \varepsilon \omega/c } \\
  2 i \sqrt{\varepsilon ^2-c^4 m^2} e^{i (l+1)\phi}
   J_{l+1}\left(\frac{2\lambda'\sqrt{ \left(\varepsilon ^2-c^4 m^2\right)}}{B }\right)-\frac{c \left(c^2 m+\varepsilon \right) \left(\dot{f}_1(\xi)+i
   \dot{f}_2(\xi)\right) e^{il\phi}J_l\left(\frac{2\lambda'\sqrt{ \left(\varepsilon ^2-c^4
   m^2\right)}}{B }\right)}{\varepsilon \omega/c } \\
\end{array}
\right),\nonumber\\
\end{align}
\end{widetext}
 with
 \begin{align*}
 F(x',y',\xi)=\frac{\mathcal{N}}{2B} \left(\frac{2c \hbar }{B}\right)^l e^{ i \left(\Phi-t \varepsilon/\hbar \right)},
 \end{align*}
 \begin{figure}
  \includegraphics[width=0.8\columnwidth]{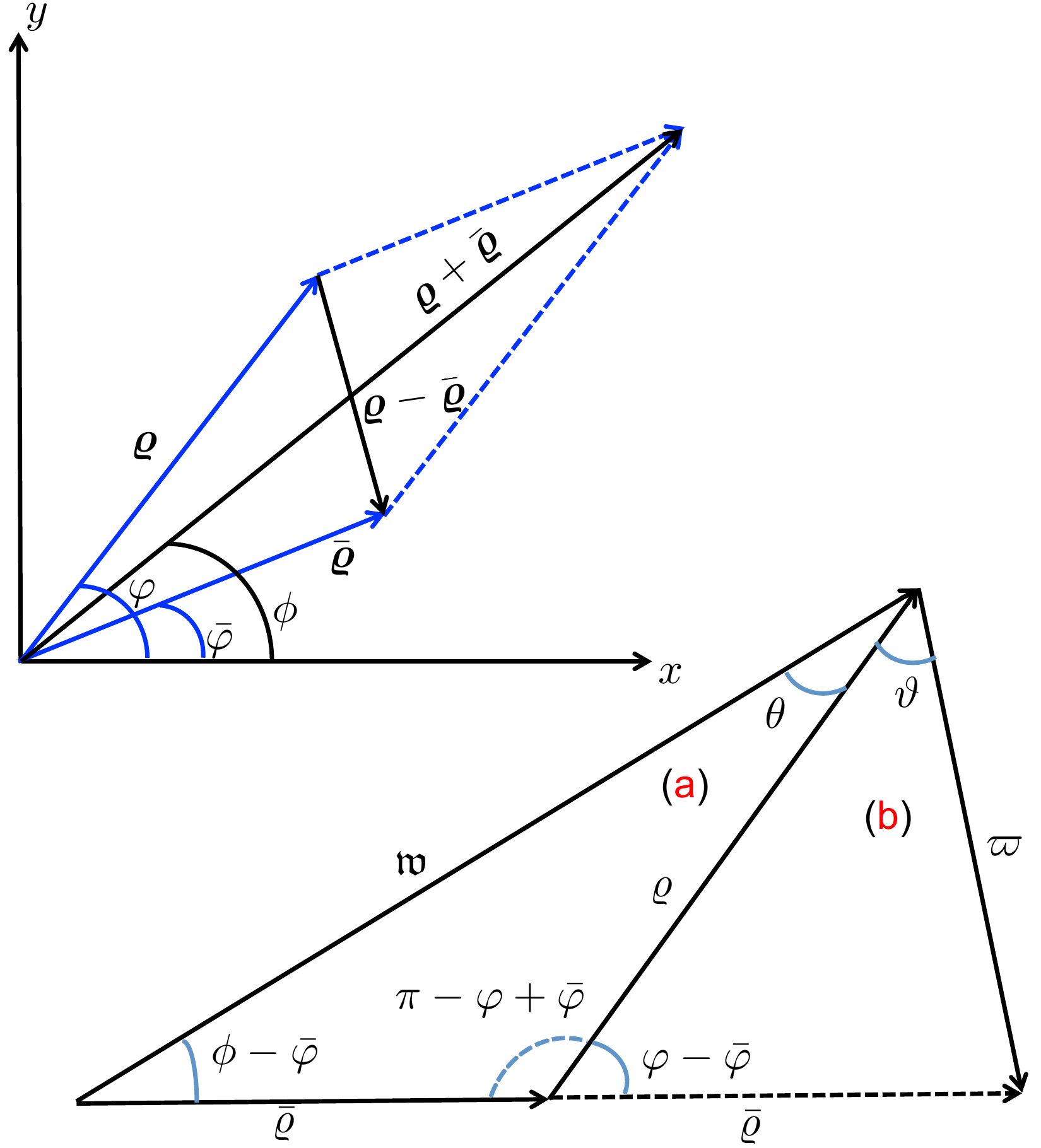}
  \caption{(colour online) Illustration of the vectors $\bar{\boldsymbol{\varrho}},\boldsymbol{\varrho}, \boldsymbol{\varrho}+\bar{\boldsymbol{\varrho}}$ and $\boldsymbol{\varrho}-\bar{\boldsymbol{\varrho}}$ along with the respective triangles formed by them.} \label{Diagrams}
\end{figure}
 and $\phi=\tan ^{-1}\left(\frac{y'}{x'}\right)$.
In order to cast the Dirac spinor (\ref{BesselW}) in a form similar to the one given in Eq. (6) of Ref. \cite{hayrapetyan2014interaction} we make the following definitions. Let $\boldsymbol{\varrho}=(x,y)$ and $\bar{\boldsymbol{\varrho}}=\frac{c^3}{\varepsilon \omega ^2 }(f_1(\xi),f_2(\xi))$. Their norms are $\varrho=\sqrt{x^2+y^2}$ and $\bar{\varrho}=\frac{c^3}{\varepsilon\omega ^2 }\sqrt{f_1(\xi)^2+f_2(\xi)^2}$. The vectors $\boldsymbol{\varrho}$ and $\bar{\boldsymbol{\varrho}}$, which are at an angle $\varphi$ and $\bar{\varphi}$ with the $x$ axis, respectively, are depicted in Fig. \ref{Diagrams}, along with the vector sum $\boldsymbol{\varrho}+\bar{\boldsymbol{\varrho}}$, which is at an angle $\phi$ with the $x$ axis; also shown are the triangles formed by the vectors $\bar{\boldsymbol{\varrho}},\boldsymbol{\varrho}, \boldsymbol{\varrho}+\bar{\boldsymbol{\varrho}}$ and $\boldsymbol{\varrho}-\bar{\boldsymbol{\varrho}}$. For the vectors displayed in the triangle (b) we have the following identity (see chap. XI of Ref. \cite{watson1995treatise})
 \begin{align}\label{BesselSum1}
 e^{-i\nu\vartheta}J_\nu(\varpi)=\sum_{m=-\infty}^\infty J_{m}(\bar{\varrho})J_{\nu+m}(\varrho)e^{-im\left(\varphi-\bar{\varphi}\right)},
 \end{align}
 where $\varpi=\sqrt{\varrho^2+\bar{\varrho}^2-2\varrho\bar{\varrho}\cos\left(\varphi-\bar{\varphi}\right)}=|\boldsymbol{\varrho}-\bar{\boldsymbol{\varrho}}|$ with
 $$
 \vartheta=\tan^{-1}\left(\frac{\bar{\varrho}\sin\left(\varphi-\bar{\varphi}\right)}{\varrho-\bar{\varrho}\cos\left(\varphi-\bar{\varphi}\right)}\right),
 $$
and for the vectors displayed in the triangle (a) we have, by making $\left(\varphi-\bar{\varphi}\right)\rightarrow\pi-\left(\varphi-\bar{\varphi}\right)$ in (\ref{BesselSum1})
   \begin{align}\label{besselSum}
 e^{-i\nu\theta}J_\nu(\mathfrak{w})=\sum_{m=-\infty}^\infty (-1)^mJ_{m}(\bar{\varrho})J_{\nu+m}(\varrho)e^{im\left(\varphi-\bar{\varphi}\right)},
 \end{align}
 where $\mathfrak{w}=\sqrt{\varrho^2+\bar{\varrho}^2+2\varrho\bar{\varrho}\cos\left(\varphi-\bar{\varphi}\right)}=|\boldsymbol{\varrho}+\bar{\boldsymbol{\varrho}}|$ and
 $$
 \theta=\tan^{-1}\left(\frac{\sin\left(\varphi-\bar{\varphi}\right)}{\varrho/\bar{\varrho}+\cos\left(\varphi-\bar{\varphi}\right)}\right).
 $$
 Since the sum of the internal angles of a triangle should be $\pi$, from triangle (a) we get $\theta=\varphi-\phi$, and (\ref{besselSum}) becomes
 $$
  e^{i\nu\phi}J_\nu(\mathfrak{w})=\sum_{m=-\infty}^\infty (-1)^mJ_{m}(\bar{\varrho})J_{\nu+m}(\varrho)e^{i(m+\nu)\varphi-im\bar{\varphi}}.
  $$
 Hence, we have the following
 \begin{align*}
& e^{il\phi}J_l\left(\frac{\sqrt{\left(x'^2+y'^2\right) \left(\varepsilon ^2-c^4 m^2\right)}}{c \hbar
   }\right)\\
   &=\sum_{m=-\infty}^\infty (-1)^mJ_{m}(k\bar{\varrho})J_{l+m}(k\varrho)e^{i(m+l)\varphi-im\bar{\varphi}},
 \end{align*}
with $k=\frac{\sqrt{\varepsilon ^2-c^4 m^2}}{c\hbar}$. It can be made equivalent to Eq. (6) of Ref. \cite{hayrapetyan2014interaction} by choosing $\bar{\varphi}=\pi/2$ which implies that $f_1(\xi)=0$, thus corresponding to a laser field linearly polarised along the $y$ axis. Thus, with RDI method we obtain straightforwardly the OAM Volkov-Bessel wave function in a more simple form than was available in \cite{hayrapetyan2014interaction}.

\subsubsection{Homogeneous magnetic field plus laser field}\label{sec5}

We apply the Lorentz transformation to the homogeneous magnetic field case, which leads to the Redmond solution. By substituting $H(\lambda)=e^{-\lambda'^2}$ in Eqs. (\ref{RFields}) and additing (\ref{VecPotTD1}) to the resulting formulas, we get
\begin{align}\label{VecPotTD3}
eA_0^r&=-\frac{B^2\left(x' \dot{f}_2(\xi)-y' \dot{f}_1(\xi)\right)}{2 \varepsilon  \omega\hbar }=eA_0^R,\nonumber\\
 eA_1^r&=-\frac{B^2 y'}{2c^2\hbar }+\frac{c  \dot{f}_1(\xi)}{ \omega }=eA_1^R+\frac{2\varepsilon}{c}\mathcal{A}_1,\nonumber\\
 eA_2^r&=\frac{B^2 x'}{2 c^2\hbar }+\frac{c  \dot{f}_2(\xi)}{ \omega }=eA_2^R+\frac{2\varepsilon}{c}\mathcal{A}_2,\nonumber\\
 eA_3^r&=eA_0^r=eA_3^R.
\end{align}
The physical meaning and origin of $A_\mu^R$ will be elucidated in Sec. \ref{subsection2C} bellow. It is noteworthy that the above vector potential obeys the Lorentz condition $\partial_\mu A^\mu=0$.
The electromagnetic fields calculated from (\ref{VecPotTD3}) are
\begin{align}
e\boldsymbol{E}&=\frac{B^2c }{\varepsilon\omega\hbar}\left( \dot{f}_2(\xi ),-\dot{f}_1(\xi ),0\right)-c\left(  \ddot{f}_1(\xi ), \ddot{f}_2(\xi ),0\right),\label{eETHB}\\
   &=e\boldsymbol{E}_R+e\boldsymbol{E}_L,\nonumber\\
   e\boldsymbol{B}&=\frac{B^2}{\varepsilon\hbar\omega}\left(\dot{f}_1(\xi ),\dot{f}_2(\xi ),0\right)+\left(  \ddot{f}_2(\xi ),- \ddot{f}_1(\xi ),0\right)+\frac{B^2 }{c^2\hbar}\hat{z},\nonumber\\
   &=e\boldsymbol{B}_R+e\boldsymbol{B}_L+\frac{B^2 }{c^2\hbar}\hat{z}\label{eBTHB}.
 \end{align}
The transformed Dirac spinor is
\begin{widetext}
\begin{align}\label{psiTH}
&\psi_T=e^{\frac{-i\varepsilon t}{\hbar}}F(x',y',\xi)\left(
\begin{array}{c}
 \frac{4 \left(c^2 m+\varepsilon \right) \, _1F_1\left(-n;\frac{2l+2}{2};2 (\lambda')
   ^2\right)}{B}-\frac{4 B c n (x'+i y') \left( \dot{f}_2(\xi)+i  \dot{f}_1(\xi)\right) \,
   _1F_1\left(1-n;\frac{2l+4}{2};2 (\lambda')^2\right)}{\varepsilon  (2l+2) \omega  \hbar } \\
 \frac{2 c^2 \left(c^2 m+\varepsilon \right) \left( \dot{f}_1(\xi)+i  \dot{f}_2(\xi)\right) \,
   _1F_1\left(-n;\frac{2l+2}{2};2 (\lambda')^2\right)}{B \varepsilon  \omega } \\
 \frac{4 B c n (y'-i x') \left( \dot{f}_1(\xi)-i  \dot{f}_2(\xi)\right) \,
   _1F_1\left(1-n;\frac{2l+4}{2};2(\lambda')^2\right)}{\varepsilon  (2l+2) \omega  \hbar } \\
 \frac{8 i B n (x'+i y') \, _1F_1\left(1-n;\frac{2l+4}{2};2 (\lambda')^2\right)}{c (2l+2) \hbar
   }-\frac{2 c^2 \left(c^2 m+\varepsilon \right) \left( \dot{f}_1(\xi)+i  \dot{f}_2(\xi)\right) \,
   _1F_1\left(-n;\frac{2l+2}{2};2 (\lambda')^2\right)}{B \varepsilon  \omega } \\
\end{array}
\right),\nonumber\\
&F(x',y',\xi)= \mathcal{N}\left(e^{i\arctan\left(\frac{ y'}{ x'}\right)}\lambda'\right)^{l} e^{\frac{-B^2 \left(x'^2+y'^2\right)}{4\hbar^2 c^2}}\exp\left(\frac{ i
   \left(2 \hbar \Phi -2 t \sqrt{2 B^2 n +c^4
   m^2}\right)}{2 \hbar }\right).
\end{align}
\end{widetext}
Moreover, the averaged component of the current in the $z$ direction is just
\begin{align}\label{aveJz}
\langle J_z\rangle=c^4\left(\frac{\dot{f}_1(\xi)^2+\dot{f}_2(\xi)^2}{2\varepsilon^2\omega^2}\right),
\end{align}
which is simply a constant for a circularly polarised sinusoidal field. Hence, in the $p_z=0$ case we recover the resonant solutions to the Redmond field configuration (see, for instance, \cite{Bagrov1968} where the solution to the classical problem of the radiation emitted by an electron in a circularly polarised plane wave field and a longitudinal homogeneous magnetic field is given).
We want to emphasise that the above solution for the homogeneous magnetic field plus laser field generalises the one discussed in Ref. \cite{van2017nonuniform},  which only considers the relativistic Landau levels. Moreover, it is
noteworthy that even though mathematically our solution coincides with Redmond's, the electromagnetic fields in our case have a different origin as the combination of radiation fields, laser field and magnetic field; in contrast the Redmond configuration is comprised of a laser field and a magnetic field.

\subsubsection{Inhomogeneous magnetic field plus laser field}\label{sec6}

We now apply the Lorentz transformation to the inhomogeneous magnetic field case, which leads to a new solution to the Dirac equation. By substituting $H(\lambda)=e^{-\lambda'/2}$ in Eqs. (\ref{RFields}) and additing (\ref{VecPotTD1}) to the resulting formulas, we get
\begin{align}\label{VecPotTD2}
eA_0^r&=-\frac{Bc \left(x' \dot{f}_2(\xi)-y' \dot{f}_1(\xi)\right)}{4 \varepsilon  \omega
   \sqrt{x'^2+y'^2}}=eA_0^R,\nonumber\\
 eA_1^r&=-\frac{B y'}{4 c \sqrt{x'^2+y'^2}}+\frac{ c\dot{f}_1(\xi)}{ \omega }=eA_1^R+\frac{2\varepsilon}{c}\mathcal{A}_1,\nonumber\\
 eA_2^r&=\frac{B x'}{4 c \sqrt{x'^2+y'^2}}+\frac{c  \dot{f}_2(\xi)}{ \omega }=eA_2^R+\frac{2\varepsilon}{c}\mathcal{A}_2,\nonumber\\
 eA_3^r&=eA_0^r=eA_3^R,
\end{align}
also obeying the Lorentz condition $\partial_\mu A^\mu=0$. Moreover, the origin of $A_\mu^R$ is the same as for the case discussed in Sec. \ref{sec5} and is explained in Sec. \ref{subsection2C} bellow.
The transformed Dirac spinor is
\begin{widetext}
\begin{align}\label{psiTIH}
\psi_T=&e^{\frac{-i\varepsilon t}{\hbar}+i\Phi}F(x',y',\xi)\left(
\begin{array}{c}
 \frac{8 \left(c^2 m+\varepsilon \right) L_n^M\left(\frac{(M+1) \lambda' }{M+2
   n+1}\right)}{B}+\frac{B c (y'-i x') \left(\dot{f}_1(\xi)-i \dot{f}_2(\xi)\right) \left((M+1)
   L_{n-1}^{M+1}\left(\frac{(M+1) \lambda' }{M+2 n+1}\right)-n L_n^M\left(\frac{(M+1) \lambda'
   }{M+2 n+1}\right)\right)}{\varepsilon  \lambda'  \omega  \hbar  (M+2 n+1)} \\
 \frac{4 c^2 \left(c^2 m+\varepsilon \right) \left(\dot{f}_1(\xi)+i  \dot{f}_2(\xi)\right)
   L_n^M\left(\frac{(M+1) \lambda' }{M+2 n+1}\right)}{B \varepsilon  \omega } \\
 \frac{c (x'+i y') \left( \dot{f}_2(\xi)+i \dot{f}_1(\xi)\right) \left(2 B n L_n^M\left(\frac{(M+1)
   \lambda' }{M+2 n+1}\right)-2 B (M+1) L_{n-1}^{M+1}\left(\frac{(M+1) \lambda' }{M+2
   n+1}\right)\right)}{\lambda'  \omega  \hbar  \sqrt{B^2 n (M+n+1)+4 c^4 m^2 (M+2 n+1)^2}} \\
 \frac{2 B \left(i (M+1) (x'+i y') L_{n-1}^{M+1}\left(\frac{(M+1) \lambda' }{M+2 n+1}\right)+n (y'-i
   x') L_n^M\left(\frac{(M+1) \lambda' }{M+2 n+1}\right)\right)}{c \lambda'  \hbar  (M+2
   n+1)}-\frac{4 c^2 \left(c^2 m+\varepsilon \right) \left(\dot{f}_1(\xi)+i  \dot{f}_2(\xi)\right)
   L_n^M\left(\frac{(M+1) \lambda' }{M+2 n+1}\right)}{B \varepsilon  \omega } \\
\end{array}
\right),\nonumber\\
&F(x',y',\xi)=\frac{\mathcal{N}}{8} (\lambda')^{M/2} \exp \left(-\frac{\lambda'(M+1)}{2 (M+2 n+1)}+\frac{1}{2} i M \tan
   ^{-1}\left(\frac{y'}{x'}\right)\right)
\end{align}
\end{widetext}
The electromagnetic fields calculated from (\ref{VecPotTD2}), as well as charge and current densities, are
\begin{widetext}
\begin{align}
e\boldsymbol{E}&=\frac{Bc^2 }{4
   \sqrt{x'^2+y'^2}}\left(\frac{ \dot{f}_2(\xi )}{\varepsilon\omega},-\frac{\dot{f}_1(\xi )}{\varepsilon\omega},0\right)-c\left(  \ddot{f}_1(\xi ), \ddot{f}_2(\xi ),0\right)=e\boldsymbol{E}_R+e\boldsymbol{E}_L,\label{eET}\\
   e\boldsymbol{B}&=\frac{Bc }{4
   \sqrt{x'^2+y'^2}}\left(\frac{ \dot{f}_1(\xi )}{ \varepsilon  \omega },\frac{\dot{f}_2(\xi )}{\varepsilon  \omega},0\right)+\left(  \ddot{f}_2(\xi ),- \ddot{f}_1(\xi ),0\right)+\frac{B }{4c
   \sqrt{x'^2+y'^2}}\hat{z}=e\boldsymbol{B}_R+e\boldsymbol{B}_L+\frac{B }{4c\sqrt{x'^2+y'^2}}\hat{z},\label{eBT}\\
   e\mu_0\rho_e&=\frac{Bc^2 \left(y' \dot{f}_1(\xi )-x' \dot{f}_2(\xi )\right)}{4\varepsilon \omega\,  \left(x'^2+y'^2\right)^{3/2}},\quad e\mu_0  \boldsymbol{J}_e=\frac{B}{4 c \left(x'^2+y'^2\right)^{3/2}}\left(-y',x',\frac{ c\left(y' \dot{f}_1(\xi )-x' \dot{f}_2(\xi )\right)}{\varepsilon
   \omega}\right).
 \end{align}
\end{widetext}
 Moreover, the averaged component of the current in the $z$ direction is just
\begin{align}\label{aveJzIB}
\langle J_z\rangle=c^4\left(\frac{\dot{f}_1(\xi)^2+\dot{f}_2(\xi)^2}{2\varepsilon^2\omega^2}\right),
\end{align}
which is  identical to Eq. (\ref{aveJz}).
 From the above charge, current and electromagnetic fields it is clear that if we put $f_1(\xi )=f_2(\xi )=0$ we recover the time independent solution of the previous section. It is noteworthy that the electromagnetic fields above have the following properties
 $$
 \boldsymbol{B}\cdot\boldsymbol{E}=0,\quad (e\boldsymbol{E})^2-(ce\boldsymbol{B})^2=-\frac{B^2}{16(x'^2+y'^2)},
 $$
the same being true for the homogeneous magnetic field plus laser case. This is expected, since the time dependent fields are a direct consequence of the Lorentz transformation (\ref{LorentzNull}) and the only field present in the time independent cases are the magnetic fields along the $z$ axis.

It should be noticed that the spin vector $\mathfrak{s}\!\!\!/_r$ is the same for  all the above solutions. Moreover the averaged velocity density of the electron for both magnetic field solutions with respect to the transverse coordinates $(x',y')$ turns out to be
\begin{align}\label{AvVH}
\langle J_\mu^r\rangle&=\langle J_\mu\rangle\nonumber\\
&+\left(2\tan^2(\theta/2),-\frac{c^2\dot{f}_1(\xi)}{\varepsilon\omega},-\frac{c^2\dot{f}_2(\xi)}{\varepsilon\omega},2\tan^2(\theta/2)\right),
\end{align}
where $\theta$ is defined in (\ref{PolarLaser}) and the average of a local observable $\mathcal{O}$ is defined as $\langle\mathcal{O}\rangle=\int_0^\infty\int_0^{2\pi}\mathcal{O} \lambda' d\lambda' d\phi$. Furthermore, the average transversal coordinates $\langle(x',y')\rangle=\int_0^\infty\int_0^{2\pi}(x',y')J_0\lambda' d\lambda' d\phi$ are
\begin{align*}
\langle(x',y')\rangle=\frac{\pi  c^3 n \hbar  (M+n+1)}{ (M+1) \omega  \varepsilon^2  (M+2 n+1)}(\dot{f}_2(\xi),-\dot{f}_1(\xi)),
\end{align*}
while for the homogeneous magnetic field case, we have
\begin{align*}
\langle(x',y')\rangle=-\frac{2 \pi  c^3 n \hbar }{ \omega  \varepsilon^2 }(\dot{f}_2(\xi),-\dot{f}_1(\xi)).
 \end{align*}
 In comparing both results we see that, despite both being similar, while the averaged transversal coordinates have the same form as in the classical resonant case for the Redmond solutions, they differ significantly from the classical case for the inhomogeneous magnetic field. This can be seen by numerically integrating the Lorentz force equation for electromagnetic fields of the same form as Eqs. (\ref{eET}) and (\ref{eBT}). This is a consequence of the fact that the Redmond solutions correspond to coherent quantum states \cite{doi:10.1002/andp.19834950102} while the inhomogeneous magnetic field states do not. In fact, given that only Gaussian states have minimum uncertainty, from the form of Eq. (\ref{MagFB}) which is also valid in the presence of the laser fields, we can make the general claim that  $H(\lambda')$ is a Gaussian if and only if $e\boldsymbol{B}$ is constant. Hence, for  all solutions to the Dirac equation corresponding to the combination of plane electromagnetic waves and a magnetic field along the wave’s propagation direction with an arbitrary perpendicular profile, the Redmond solution stands out as being the \textit{only} one corresponding to a relativistic  coherent state.

\subsection{Physical meaning of the time dependent electromagnetic fields}\label{subsection2C}

Let us now investigate the source of the electric (\ref{eETHB}), (\ref{eET}) and magnetic (\ref{eBT}), (\ref{eBTHB}) fields. Our reasoning is similar to the one used in chapter 10 of Ref. \cite{BaylisBook1996}. They are separated as ``radiation fields" with subscript R and laser fields with subscript L, along with the original magnetic fields (the inhomogeneous one in terms of the transformed coordinates $x'$ and $y'$) which comes from the Lorentz transformation (\ref{LorentzNull}) applied to the matrix spinor $\Psi$. The radiation fields are given by the motion induced by the laser fields of the original static current distribution that generates the homogeneous and inhomogeneous magnetic fields (given that homogeneous magnetic fields can also be produced by solenoidal currents which are assumed to be far from the electron). From (\ref{RFields}) we recover the ``radiation" vector potentials (\ref{VecPotTD3}) and (\ref{VecPotTD2}) given by the respective choices of the function $H(\lambda')$. For instance, by choosing $H(\lambda')=e^{-\lambda'/2}$, we see at once that the transformed vector potential (\ref{RFields}) leads to the radiation fields (\ref{eET}) and (\ref{eBT}) as well as the transformed inhomogeneous magnetic field along the $\hat{z}$ axis. The same is true for the homogeneous magnetic field case as can be seen by replacing  $H(\lambda')=e^{-\lambda'^2}$ in  (\ref{RFields}).%\\

\section{Conclusion and Outlook}\label{section3}

We have shown how applying RDI allows to generate solutions of the Dirac equation employing Hestenes' geometrical representation and the physical intuition behind the geometry. We have illustrated RDI method to construct the complete set of eigenvalues and eigenfunctions for an electron in a homogeneous as well as an inhomogeneous magnetic field.
Even though those solutions were already discovered a long time ago, our new derivation shed light on their geometrical
properties. Moreover, RDI also provided a rather straightforward way to generalise those solutions for the case in which
a plane wave electromagnetic field is added, leading to the discovery of a novel solution to the Dirac equation. Furthermore, we show that RDI also can give the self-consistent electromagnetic fields generated by the electron dynamics
when the plane wave field is included.

Having further illustrated the potential of RDI, some important points remain to be explored. For instance, what is the role of the parameter $\beta$ in the
solutions to the Dirac equation, given that it is connected with quantum states containing particle and antiparticle admixtures. Here we present our conjecture that $\beta$ is necessary (albeit not suficient) for the probability density $J_0$ to have compact support, a property that requires negative energy components in the wave packet expansion. On the other hand, we can construct square integrable, although without compact support, wave functions composed of only positive energy states. Understanding this feature would make it possible for us to separate those types of external fields that allows for a clear separation between positive and negative energy states (this is the case for  all the solutions presented here) from those which don't. Thus allowing an effective control of wave packet spreading in external fields, having important implications, for instance, in the attoscience, where recollisions play a pivotal role. This may help to control and optimize high-order harmonic generation from electron-atom rescattering in the relativistic regime \cite{HHG}, and to permit the laser-driven high-energy collider \cite{collider}.

Finally, it is noteworthy to call attention to an important discussion concerning the Hestenes formulation of the Dirac equation, which has been hitherto overlooked in the literature, that it seems to correspond to a straightforward relativistic generalisation \cite{DORAN1996271} of the de Broglie-Bohm pilot wave theory \cite{PhysRev.85.166}. RDI further highlights such correspondence since upon identifying the particle trajectories with the Dirac's fluid streamlines, well defined solutions of the Dirac equation with clear physical and geometrical meanings are derived. From this point of view there is also the possibility of extending RDI to account for other types of interactions besides electromagnetism, simply by relaxing the constraints (\ref{constraints}) imposed on the form of $A\!\!\!/$ in a simpler and more transparent way compared with the standard Dirac approach.

\section*{acknowledgement}

A.G.C. acknowledges financial support from the Humboldt Research Fellowship for Postdoctoral Researchers and insightful discussions with Renan Cabrera.

\appendix

\section{Finite Lorentz transformations}\label{Appendix1}

A finite Lorentz transformation can be written in terms of the exponential of products of gamma matrices as follows
\begin{align}\label{PsiExpMap}
\mathcal{R}=e^{\frac{1}{2}\gamma_\mu\gamma_\nu\omega^{\mu\nu}},
\end{align}
where
$$
\omega^{\mu\nu}=\begin{pmatrix}
   0 & a^1 &a^2   & a^3 \\
   -a^1   &   0 & -b^3   & b^2 \\
   -a^2   &  b^3  & 0   & -b^1 \\
   -a^3   &  -b^2   & b^1  &  0
    \end{pmatrix},
$$
is a real, antisymmetric matrix. The above product of gamma matrices satisfy the following commutation relations
$$
\frac{1}{2}[\gamma_\mu\gamma_\nu,\gamma_\rho\gamma_\tau]=\gamma_\mu\gamma_\tau\boldsymbol{\eta}_{\nu,\rho}-\gamma_\mu\gamma_\rho\boldsymbol{\eta}_{\nu,\tau}-\gamma_\nu\gamma_\tau\boldsymbol{\eta}_{\mu,\rho}+\gamma_\nu\gamma_\rho\boldsymbol{\eta}_{\mu,\tau}.
$$
These commutation relations defines an algebra over a $6$-dimensional space. Hence it forms a vector space whose basis is given by $\gamma_\mu\gamma_\nu$. Such a basis provides a convenient representation of the Lie group $SL(2,\mathbb{C})$ by $4\times4$ complex matrices. From this it follows that the elements of the matrix $\omega^{\mu\nu}$ are the coordinates of a given point in the Lie group manifold. The exponent of (\ref{PsiExpMap}) have the following explicit form
\begin{align}\label{groupPar}
\frac{1}{2}\gamma_\mu\gamma_\nu\omega^{\mu\nu}=a^k\alpha_k-b^k\boldsymbol{i}\alpha_k.
\end{align}
Thus, from the properties of the Dirac matrices, the matrix exponential can always be written in the polar form
\begin{align}
e^{\frac{1}{2}\gamma_\mu\gamma_\nu\omega^{\mu\nu}}&=e^{\frac{w}{2}\alpha_k\tilde{a}^k}e^{\frac{\theta}{2}\boldsymbol{i}\alpha_k\tilde{b}^k},\quad \tilde{a}^k=\frac{a^k}{\sqrt{a_1^2+a_2^2+a_3^2}},\nonumber\\
\tilde{b}^k&=\frac{b^k}{\sqrt{b_1^2+b_2^2+b_3^2}},
\end{align}
with $w$, $\theta$ real, $e^{\frac{w}{2}\alpha_k\tilde{a}^k}$ is a Hermitian matrix representing a boost and $e^{\frac{\theta}{2}\boldsymbol{i}\alpha_k\tilde{b}^k}$ is a unitary matrix representing a spatial rotation.
\section{Spin vectors}\label{Spinors}
\begin{figure}
  \includegraphics[width=0.5\textwidth]{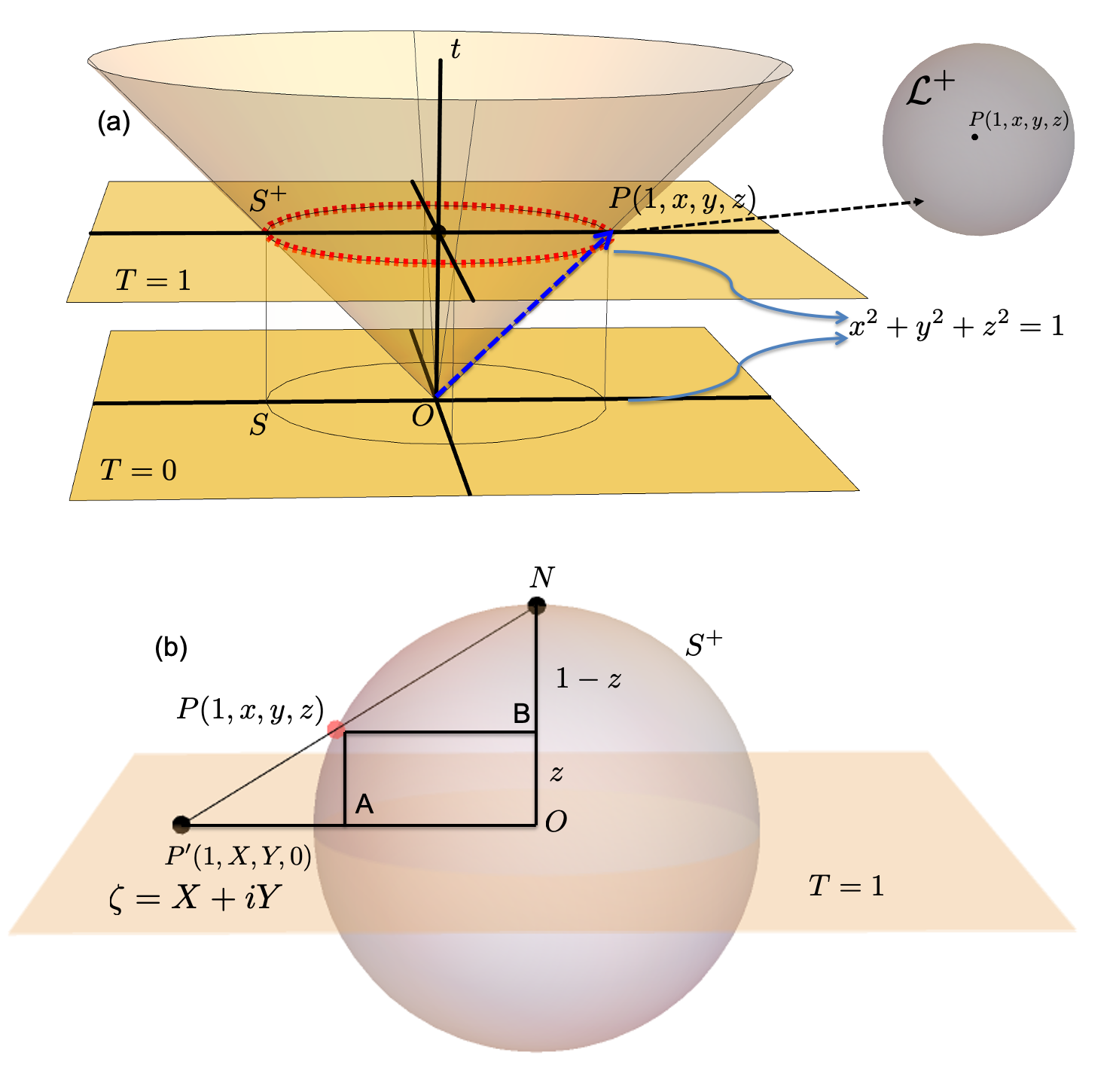}
  \caption{(a) The $\mathcal{L}^+$ sphere corresponds to the intersection between the plane and the light cone. (b) Stereographic projection from the sphere onto the complex plane.}
     \label{Stereo3}
\end{figure}

In order for this paper to be self-contained, here we give a geometrical description of spin vectors. We follow the discussion given in chap. 2 of Ref. \cite{penrose1984spinors}. Coordinates will be assigned to the light cone (i.e., the set of null vectors whose vertex lies in the particle's trajectory as illustrated in Fig. \ref{ManifoldCurves}) in Minkowski spacetime using complex numbers. From this an insightful geometrical interpretation of spinors will be reached. Given a Minkowski tetrad $(\boldsymbol{e}_0,\boldsymbol{e}_1,\boldsymbol{e}_2,\boldsymbol{e}_3)$, \textit{any} vector in spacetime can be written as
$$
U=T\boldsymbol{e}_0 +X\boldsymbol{e}_1 +Y\boldsymbol{e}_2 +Z\boldsymbol{e}_3.
$$
In the case of null vectors, we have
\begin{align}\label{LightCone}
T^2-X^2-Y^2-Z^2=0.
\end{align}
The light cone is the set of null directions in spacetime passing through the origin $O$ as shown in Fig. \ref{Stereo3}. In this respect, the two vectors $\pm U$ are considered to have opposite directions along the light cone, $+U$ is future oriented (i.e., lying above the plane through the origin in Fig. \ref{Stereo3}) while $-U$ is past-oriented (i.e., lying bellow the plane through the origin in Fig. \ref{Stereo3}).
As depicted in Fig. \ref{Stereo3}, the abstract space whose points lie in the future light cone is denoted by the sphere $\mathcal{L}^+$, which can be represented in any given coordinates $(T,X,Y,Z)$ by the intersection $S^+$ of the future light cone (\ref{LightCone}) with the hyperplane $T=1$. Such intersection is given by the sphere $x^2+y^2+z^2=1$ (this sphere is depicted in Figs. \ref{Stereo}, \ref{Stereo2} and \ref{Stereo3}). Any point in the hyperplane $T=1$ is represented by the homogeneous coordinates $(X/T,Y/T,Z/T)$. The interior and exterior points of $S^+$ corresponds to time-like and space-like future directions, respectively. Moreover, any null direction along the cone is represented as a single point on the sphere $\mathcal{L}^+$.

A complex number $\zeta$ can be assigned to any given point $(x,y,z)$ on the sphere via the stereographic projection of a sphere onto a plane (see Fig.\ref{Stereo3}). From the fact that the triangles $NCP'$ and  $NOP$ are similar, we have the following
\begin{align}\label{stereographic}
X&=\frac{x}{1-z},\quad Y=\frac{y}{1-z},\nonumber\\
x&=\frac{1}{\zeta\zeta^\ast+1}(\zeta+\zeta^\ast),\quad y=\frac{1}{i(\zeta\zeta^\ast+1)}(\zeta-\zeta^\ast),\quad z=\frac{\zeta\zeta^\ast-1}{\zeta\zeta^\ast+1}.
\end{align}
where $\ast$ corresponds to complex conjugation. Hence, the sphere $S^+$ is the Riemann sphere of the complex plane, a well-known representation of the complex numbers including the point at infinity corresponding to the north pole $N=P(1,0,0,1)$. A more convenient way to include the point at infinite is to label the points of $S^+$ by two complex numbers instead of one. That is,
$$
\zeta=\chi/\eta,
$$
which can be written as a column vector $\phi=(\chi,\eta)^T$, the sought after \textit{spin vector}.
These are the homogeneous complex coordinates, so that the pairs $(\chi,\eta)$ and $\Lambda(\chi,\eta)$ represent the same point on $S^+$ for any complex number $\Lambda\neq0$. With these coordinates the point $\zeta=\infty$ is represented by $(1,0)$.
In terms of the homogeneous coordinates, the equations (\ref{stereographic}) become
\begin{align}\label{stereographic2}
&x=\frac{1}{\chi\chi^\ast+\eta\eta^\ast}(\chi\eta^\ast+\chi^\ast\eta),\quad y=\frac{1}{i(\chi\chi^\ast+\eta\eta^\ast)}(\chi\eta^\ast-\chi^\ast\eta),\nonumber\\
&z=\frac{\chi\chi^\ast-\eta\eta^\ast}{\chi\chi^\ast+\eta\eta^\ast}.
\end{align}

Consider, then, a complex linear (non-singular) transformation of $\chi$ and $\eta$
\begin{align}\label{LorentzT}
\chi\rightarrow\chi'=\alpha\chi+\beta\eta,\nonumber\\
\eta\rightarrow\eta'=\gamma\chi+\delta\eta
\end{align}
where $\alpha$, $\beta$, $\gamma$ and $\delta$ are arbitrary complex numbers subject only to the condition $\alpha\delta-\beta\gamma\neq0$ (non-singularity). Expressed in terms of $\zeta$, the transformation (\ref{LorentzT}) becomes
\begin{align}\label{Mobius}
\zeta\rightarrow\zeta'=\frac{\alpha\zeta+\beta}{\gamma\zeta+\delta}.
\end{align}
The transformation (\ref{LorentzT}) (or (\ref{Mobius})) is called spin transformations in the context where $\zeta$ is related to the Minkowski null vector through equations (\ref{stereographic2}). In the same context, we define the spin matrix $A$ by
\begin{align}\label{SpinMatrix}
A = \begin{pmatrix}
    \alpha  & \beta \\
    \gamma    &  \delta
    \end{pmatrix},\quad \mbox{det}({A})\neq0.
\end{align}
In terms of $A$, (\ref{LorentzT}) takes the form
\begin{align}\label{ATransf}
\begin{pmatrix}
        \chi' \\ \eta'
       \end{pmatrix}=A\begin{pmatrix}
        \chi \\ \eta
       \end{pmatrix}.
\end{align}
This is regarded as the spin transformation matrix in a 2-dimensional complex space with the basic spin vector
\begin{align}\label{2spinor}
\phi=\begin{pmatrix}
        \chi \\ \eta
       \end{pmatrix},\quad \phi\rightarrow A\phi,\quad \phi^\dagger\rightarrow\phi^\dagger A^\dagger
\end{align}
with $ \phi^\dagger\phi=|\chi|^2+|\eta|^2$ an invariant up to a scaling factor under the action of the spin transformation matrix $A$. We see from (\ref{2spinor}) that $\phi$ and $\phi^\dagger$ transform in different ways but we can show that $(\chi,\eta)^T$ and $(-\eta^\ast,\chi^\ast)^T$ transform in the same way under $A$. Comparing (\ref{LorentzT}) and  (\ref{2spinor}) we have
\begin{align*}
(-\eta^\ast)'&=\delta^\ast(-\eta^\ast)-\gamma^\ast\chi^\ast,\\
(\chi^\ast)'&=-\beta^\ast(- \eta^\ast)+\alpha^\ast\chi^\ast.
\end{align*}
Hence the conjugated spin vector
\begin{align}\label{ParityT}
\begin{pmatrix}
       - \eta^\ast \\ \chi^\ast
       \end{pmatrix}=\begin{pmatrix}
    0 & -1\\
    1 & 0
    \end{pmatrix}\begin{pmatrix}
        \chi^\ast \\ \eta^\ast
       \end{pmatrix}=-i\sigma_2\phi^\ast=\dot{\phi},
\end{align}
where $\sigma_2$ is a Pauli matrix, transform under
\begin{align}\label{Conjugation}
A\rightarrow-\begin{pmatrix}
    0 & -1\\
    1 & 0
    \end{pmatrix}A^\ast \begin{pmatrix}
    0 & -1\\
    1 & 0
    \end{pmatrix}.
\end{align}
The transformation (\ref{ParityT}) acting on the spinor $\phi$ have the effect of sending the point $P$ into its antipodal point. Therefore, it is a parity transformation;
its effect on $\zeta$ is
$$
\zeta'=-\frac{1}{\zeta^\ast}.
$$
Thus, the point at infinity $\zeta'=\infty$ is then represented by $(0,1)$, that is, the south pole of the Riemann sphere.

We see from (\ref{ATransf}) that the composition of two successive spin transformations is again a spin transformation: the spin matrix of the composition is given by the product of the spin matrices of the factors. Moreover, any spin matrix has an inverse
$$
A^{-1} = \frac{1}{\mbox{det}({A})}\begin{pmatrix}
    \delta  & -\beta \\
   - \gamma    &  \alpha
    \end{pmatrix},
$$
which is also a spin matrix. Thus, the spin transformations form a group, referred to as $SL(2,\mathbb{C})$ if $\mbox{det}{(A)}=1$.

Recall that the role of the point $P(1,x,y,z)$ on $S^+$ was simply to represent a future null direction at $O$, that is, the dotted blue vector $\vec{OP}$ in Fig. \ref{Stereo3}. We could also choose any other point along the line $OP$ to represent the same null direction. In particular we could choose a point, say $L$, on $OP$ whose coordinates $(T,X,Y,Z)$ are obtained from those of $P$ by multiplying Eqs. (\ref{stereographic2}) by the factor $(\chi\chi^\ast+\eta\eta^\ast)/\sqrt{2}$. Then, the vector $\vec{OL}$ has coordinates
\begin{align}\label{stereographic3}
&T=\frac{\chi\chi^\ast+\eta\eta^\ast}{\sqrt{2}},\quad X=\frac{1}{\sqrt{2}}(\chi\eta^\ast+\chi^\ast\eta),\quad Y=\frac{1}{i\sqrt{2}}(\chi\eta^\ast-\chi^\ast\eta),\nonumber\\
&Z=\frac{\chi\chi^\ast-\eta\eta^\ast}{\sqrt{2}},
\end{align}
which can be inverted and re-expressed as
\begin{align}\label{MSpin}
\frac{1}{\sqrt{2}}\begin{pmatrix}
    T+Z & X-iY\\
    X+iY & T-Z
    \end{pmatrix}=\phi\phi^\dagger.
\end{align}
where $\dagger$ stands for Hermitian conjugation. From (\ref{ATransf}) it then follows that
\begin{align}\label{ATransf3}
&\begin{pmatrix}
    T+Z & X-iY\\
    X+iY & T-Z
    \end{pmatrix}\rightarrow\begin{pmatrix}
    T'+Z' & X'-iY'\\
    X'+iY' & T'-Z'
    \end{pmatrix}\nonumber\\
   &= A\begin{pmatrix}
    T+Z & X-iY\\
    X+iY & T-Z
    \end{pmatrix}A^\dagger.
\end{align}
This is a linear transformation of $(T,X,Y,Z)$, It is real and preserves the Minkowski norm of $U$, that is $T'^2-X'^2-Y'^2-Z'^2=T^2-X^2-Y^2-Z^2$ even if $U$ is not null. Thus, (\ref{ATransf3}) defines a \textit{Lorentz transformation}.

\subsection{$SL(2,\mathbb{C})$ and Lorentz transformations}
The spin transformation matrix $A$ with unit determinant can be represented by the set of complex $2\times2$ matrices forming a group with respect to matrix multiplication, the $SL(2,\mathbb{C})$. It also forms a vector space under multiplication by real numbers as well as matrix addition. This vector space is $4$-dimensional because the Pauli matrices $\sigma_0=I$, $\sigma_i$, $i=1,2,3$, having the property $\sigma_1\sigma_2\sigma_3=iI$ where $i=\sqrt{-1}$, forms a basis of it.

The Pauli matrices will now be a representation of the Minkowski tetrad, i.e.,
$$
(\boldsymbol{e}_0,\boldsymbol{e}_1,\boldsymbol{e}_2,\boldsymbol{e}_3)\rightarrow(\sigma_0,\sigma_1,\sigma_2,\sigma_3).
$$
Hence, we can rewrite (\ref{MSpin}) within $SL(2,\mathbb{C})$ as
\begin{align}\label{MSpinSLC}
\phi\phi^\dagger=\sigma(U)=\frac{1}{\sqrt{2}}\left(T\sigma_0+X\sigma_1+Y\sigma_2+Z\sigma_3\right)
\end{align}
The association (\ref{MSpinSLC}) defines an isomorphism of $\mathbb{R}^4$ onto $SL(2,\mathbb{C})$.
and the spin transformation takes the form
$$
\sigma(U')=A\sigma(U)A^\dagger,\quad A\in SL(2,\mathbb{C}).
$$
\comm{
\subsubsection{A representation of boosts}
Let $H\in SL(2,\mathbb{C})$ be a Hermitian matrix. The ansatz
\begin{align}\label{BoostRep}
H=\cosh(\omega/2)\sigma_0+\sinh(\omega/2)n^k\sigma_k=e^{\frac{\omega}{2}\boldsymbol{n}\cdot\vec{\sigma}}, \quad \omega\in\mathbb{R},\quad |\boldsymbol{n}|=1,
\end{align}
automatically satisfies det $H\,=1$. The matrix elements of $\Lambda_H$ can be read off from Eq. (\ref{components}). Comparing with the matrix (\ref{BoostMatrix}), we see that Eq. (\ref{BoostRep}) is a boost with velocity
$$
\boldsymbol{v}=\tanh(\omega)\boldsymbol{n}.
$$
\subsubsection{A representation of rotations}
For any unitary matrix $U\in SU(2)$ (the group of complex $2\times2$ unitary matrices with unit determinant) we may write
\begin{align}\label{RotationU}
U=\cos(\phi/2)-i\sin(\phi/2)\boldsymbol{n}\cdot\vec{\sigma}=e^{-i\frac{\phi}{2}\boldsymbol{n}\cdot\vec{\sigma}},\quad\phi\in[0,4\pi),\quad |\boldsymbol{n}|=1.
\end{align}
The matrix elements of $\Lambda_U$ are again determined by Eq. (\ref{components}). Comparison with (\ref{RotationMatrix}) shows that $\Lambda_U$ is a rotation around the axis $\boldsymbol{n}$ through an angle $\phi$. The decomposition $A=HU\in SL(2,\mathbb{C})$ corresponds to the decomposition (\ref{GroupDec}) of a proper Lorentz transformation into boosts and rotations.
}
Let us consider now the isomorphism $U\rightarrow\sigma'(U)$
\begin{align}\label{MSpinSLC2}
\sigma'(U)=\frac{1}{\sqrt{2}}\left(T\sigma_0-X\sigma_1-Y\sigma_2-Z\sigma_3\right).
\end{align}
The relation
$$
\sigma'(U)\rightarrow\sigma'(U')=B\sigma'(U)B^\dagger,\quad B\in SL(2,\mathbb{C}),
$$
also defines a Lorentz transformation, the matrices $A$ and $B$ corresponding to the same Lorentz transformation. They are connected by $B=\sigma_2(A^\dagger)^T\sigma_2=(A^\dagger)^{-1}$, which coincides with the transformation (\ref{Conjugation}). The map $A\rightarrow B$ is a group automorphism of $SL(2,\mathbb{C})$. It corresponds to the space reflection given by (\ref{ParityT}).
Since there's no element corresponding to complex conjugation in $SL(2,\mathbb{C})$, the representations $A$ and $B$ are not equivalent. Considering that a space reflection is also a Lorentz transformation, the automorphism $A\leftrightarrow(A^\dagger)^{-1}$ suggests to double the dimension of the representation space in order to obtain a matrix representation of a space reflexion. In the linear space $\mathbb{C}^4$, the matrices $A$ and $(A^\dagger)^{-1}$ are combined into $4\times4$ matrices of the form
\begin{align}\label{LorentzRotorL}
\boldsymbol{L}_A=\begin{pmatrix}
    A & 0\\
  0   & (A^\dagger)^{-1}
    \end{pmatrix},\quad A\in SL(2,\mathbb{C}).
\end{align}
The automorphism $A\leftrightarrow(A^\dagger)^{-1}$ can now be represented by the matrix
$$
\boldsymbol{L}_P=\begin{pmatrix}
    0 & I\\
   I & 0
    \end{pmatrix}.
$$
The matrices $\boldsymbol{L}_A$ forms a group under matrix multiplication which is isomorphic to $SL(2,\mathbb{C})$. The mapping $A\rightarrow\boldsymbol{L}_A$ (\ref{LorentzRotorL}) defines an injective representation of $SL(2,\mathbb{C})$ which is reducible. There are only two invariant subspaces which are mapped into each other by the matrix $\boldsymbol{L}_P$. Hence, the matrix group $\tilde{\mathfrak{L}}=\{\boldsymbol{L}_A,\boldsymbol{L}_P\boldsymbol{L}_A| \,A\in SL(2,\mathbb{C})\}$ acts irreducible on $\mathbb{C}^4$.

In order to describe the connection of $\boldsymbol{L}_A$ with Lorentz transformations more precisely, we define a suitable $4\times4$ matrix $\gamma(x)$ for each $x\in\mathbb{R}^4$, such that Lorentz transformations can be described as a similarity transformation of $\gamma(x)$. A particular choice is
\begin{align}\label{gammaM}
\gamma(x)=\begin{pmatrix}
    0 & \sigma(x)\\
 \sigma'(x)   & 0
    \end{pmatrix}.
\end{align}
Thus we obtain, with $B=(A^\dagger)^{-1}$
\begin{align}
\boldsymbol{L}_A\gamma(x)\boldsymbol{L}_A^{-1}&=\begin{pmatrix}
    0 & A\sigma(x)A^\dagger\\
B \sigma'(x)B^\dagger   & 0
    \end{pmatrix}=\gamma(\Lambda_Ax),\nonumber\\
\boldsymbol{L}_P\gamma(x)\boldsymbol{L}_P^{-1}&= \begin{pmatrix}
    0 & \sigma'(x)\\
 \sigma(x)   & 0
    \end{pmatrix}=\gamma(Px),\quad P=\begin{pmatrix}
    1 & \boldsymbol{0}^T\\
    \boldsymbol{0}   &  -\boldsymbol{1}_3
    \end{pmatrix},
\end{align}
where $\boldsymbol{0}^T=(0,0,0)$ and $\boldsymbol{1}_3$ is the $3\times3$ unit matrix; $\Lambda_A$ is defined bellow. The bijective map $\gamma:x\leftrightarrow\gamma(x)$, $x\in\mathbb{R}^4$ is an isomorphism of the vector space $\mathbb{R}^4$ and the $4$-dimensional real vector space of matrices of the form (\ref{gammaM}). The canonical basis $\{\boldsymbol{e}_0,\boldsymbol{e}_1,\boldsymbol{e}_2,\boldsymbol{e}_3\}\in\mathbb{R}^4$ is mapped to the gamma matrices
\begin{align}\label{gammaM2}
&\gamma_0=\gamma(\boldsymbol{e}_0)=\begin{pmatrix}
    0 &I\\
 I   & 0
    \end{pmatrix},\quad \gamma_k=\gamma(\boldsymbol{e}_k)=\begin{pmatrix}
    0 & \sigma_k\\
 -\sigma_k   & 0
    \end{pmatrix},\nonumber\\
    &\gamma(x)=\langle\gamma,x\rangle=\gamma^0x^0-\gamma^1x^1-\gamma^2x^2-\gamma^3x^3,
\end{align}
where $\gamma_0=\gamma^0=(\gamma^0)^\dagger, \gamma_k=-\gamma^k=(\gamma^k)^\dagger$. From the properties of the Pauli matrices it is easy to show that
\begin{align}\label{antiComm}
\frac{1}{2}\left(\gamma_\mu\gamma_\nu+\gamma_\nu\gamma_\mu\right)=\boldsymbol{\eta}_{\mu,\nu}\boldsymbol{1}.
\end{align}
By construction there is a $\boldsymbol{L}\in\tilde{\mathfrak{L}}$ for every Lorentz transformation $\Lambda_{\boldsymbol{L}}\in\mathfrak{L}$, where $\mathfrak{L}=SO(1,3)$ is the Lorentz group, such that
$$
\boldsymbol{L}\langle\gamma,x\rangle\boldsymbol{L}^{-1}=\langle \gamma,\Lambda_{\boldsymbol{L}}x\rangle,
$$
and the map $\boldsymbol{L}\rightarrow\Lambda_{\boldsymbol{L}}$ is onto and a homomorphism; it is the double cover of the Lorentz group.
We can now give a formal justification for the description given in the Appendix \ref{Appendix1}. Using $\gamma$ matrices, the matrices $\boldsymbol{L}$ corresponding to boosts and rotations become
\begin{align}
\boldsymbol{L}_B&=\mathcal{B}=e^{-\frac{\omega}{2}n_k\gamma^k\gamma^0},\quad \mbox{ boost with velocity }\boldsymbol{v}=\tanh(\omega)\boldsymbol{n},\\
\boldsymbol{L}_U&=U=e^{\frac{\phi}{2}\boldsymbol{i}n_k\gamma^k\gamma^0}, \quad \mbox{ rotation through an angle $\phi$ around $\boldsymbol{n}$},\\
\boldsymbol{L}_P&=\gamma^0,\quad\mbox{ space reflection}.
\end{align}
It can be shown that $\boldsymbol{L}^{-1}=\gamma_0\boldsymbol{L}^\dagger\gamma_0$ \cite{greiner1990relativistic,thaller2013dirac}.

\subsection{Description of the null rotation generating the laser field}

Given the results of Sec. \ref{subsection2B}, let us investigate the spinor representation of a null rotation. Since it corresponds to a translation in the complex plane, it is given by
\begin{align}\label{NullR}
\zeta'=\zeta+a,
\end{align}
where $a$ is some complex number. A spin transformation (up to a sign) generating (\ref{NullR}) is then
\begin{align}\label{NullRSpin}
\begin{pmatrix}
       \xi' \\ \eta'
       \end{pmatrix}=\begin{pmatrix}
    1& a\\
    1 & 0
    \end{pmatrix}\begin{pmatrix}
       \xi \\ \eta
       \end{pmatrix},
\end{align}
while the conjugated spin transformation, according to (\ref{ParityT}) is
\begin{align}\label{NullRSpinC}
\begin{pmatrix}
       -(\eta')^\ast \\ (\xi')^\ast
       \end{pmatrix}=\begin{pmatrix}
    1& 0\\
    -a^\ast & 1
    \end{pmatrix}\begin{pmatrix}
       -\eta^\ast \\ \xi^\ast
       \end{pmatrix}.
\end{align}
Applying the definition (\ref{LorentzRotorL}) to the spin transformation matrix in (\ref{NullRSpinC}) we arrive at
\begin{align}\label{NullR4}
\boldsymbol{L}_A=\begin{pmatrix}
    1&0&0 & 0\\
    -a^\ast&1&0&0\\
    0&0&1&a\\
  0   &0&0& 1
    \end{pmatrix}.
\end{align}
for $a=-c^2(\dot{f}_1(\xi)-i\dot{f}_2(\xi))/(\varepsilon\omega)$ it is easy to see that
$$
\boldsymbol{L}_A=T\Phi^rT^\dagger,\quad T=\frac{1}{\sqrt{2}}\left(1+\gamma_5\gamma_0\right),\quad \Phi^r=e^{\,k\!\!\!/\wedge \mathcal{A}\!\!\!/}.
$$
\bibliography{bib-proposal}

\end{document}